\documentclass{article}

\usepackage{amsmath}
\usepackage{booktabs}
\usepackage{multirow}
\usepackage{float}
\usepackage{gensymb}
\usepackage[caption=false]{subfig}
\usepackage{graphicx}
\usepackage{algpseudocode,algorithm}

\usepackage{PRIMEarxiv}

\usepackage[utf8]{inputenc} 
\usepackage[T1]{fontenc}    
\usepackage{hyperref}       
\usepackage{url}            
\usepackage{booktabs}       
\usepackage{amsfonts}       
\usepackage{nicefrac}       
\usepackage{microtype}      
\usepackage{lipsum}
\usepackage{fancyhdr}       
\usepackage{graphicx}       
\graphicspath{{media/}}     
\usepackage{natbib}
\title{Cryo-shift: Reducing domain shift in cryo-electron subtomograms with unsupervised domain adaptation and randomization}
\author
{
	Hmrishav Bandyopadhyay\\
	Department of Electronics and \\ Telecommunication Engineering \\
	Jadavpur University \\
	\And
	Zihao Deng\\
	Computational Biology Department \\
	Carnegie Mellon University \\
	\And
	Leiting Ding\\
	Computational Biology Department \\
	Carnegie Mellon University \\
	\AND
	Sinuo Liu \\
	Computational Biology Department \\
	Carnegie Mellon University \\
	\And
	Mostofa Rafid Uddin\\
	Computational Biology Department \\
	Carnegie Mellon University \\
	\And
	Xiangrui Zeng\\
	Computational Biology Department\\
	Carnegie Mellon University \\
	\And  
	Sima Behpour\\
	Computational Biology Department\\
	Carnegie Mellon University \\
	\And
	Min Xu \footnote{Corresponding Author: \texttt{mxu1@cs.cmu.edu}}\\
	Computational Biology Department\\
	Carnegie Mellon University \\
	
}
\begin{document}
	\maketitle

\begin{abstract}
	\textbf{Motivation:} 
	Cryo-Electron Tomography (cryo-ET) is a 3D imaging technology that enables the visualization of subcellular structures \textit{in situ} at near-atomic resolution. Cellular cryo-ET images help in resolving the structures of macromolecules and determining their spatial relationship in a single cell, which has broad significance in cell and structural biology. Subtomogram classification and recognition constitute a primary step in the systematic recovery of these macromolecular structures.  Supervised deep learning methods have been proven to be highly accurate and efficient for subtomogram classification, but suffer from limited applicability due to scarcity of annotated data. While generating simulated data for training supervised models is a potential solution, a sizeable difference in the image intensity distribution in generated data as compared to real experimental data will cause the trained models to perform poorly in predicting classes on real subtomograms.
	\\
	\textbf{Results:} 
	In this work, we present \textit{Cryo-Shift}, a fully unsupervised domain adaptation and randomization framework for deep learning-based cross-domain subtomogram classification. We use unsupervised multi-adversarial domain adaption to reduce the domain shift between features of simulated and experimental data. We develop a network-driven domain randomization procedure with `warp' modules to alter the simulated data and help the classifier generalize better on experimental data. We do not use any labeled experimental data to train our model, whereas some of the existing alternative approaches require labeled experimental samples for cross-domain classification.  Nevertheless, \textit{Cryo-Shift} outperforms the existing alternative approaches in cross-domain subtomogram classification in extensive evaluation studies demonstrated herein using both simulated and experimental data. 
	\\
	\textbf{Availability:} https://github.com/xulabs/aitom \\
	\textbf{Contact:} \href{mxu1@cs.cmu.edu}{mxu1@cs.cmu.edu}\\
	
\end{abstract}

	\section{Introduction}
		\begin{figure}
    	\centering
    	\includegraphics[width=0.4\columnwidth]{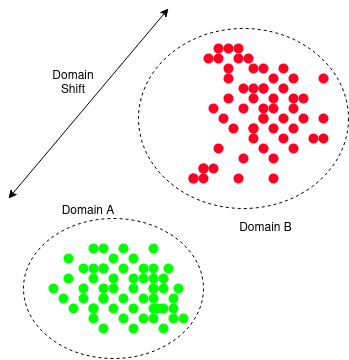}
    	\label{fig:domain}
    	\caption{Domain shift illustrated between source and target distributions}
    \end{figure}
    
	Complex macromolecules participate in a large number of biochemical processes which helps them sustain their cellular environment and governs their cellular activities. For an accurate analysis of cellular processes enabled by the interaction of these macromolecules, an in-depth analysis of their spatial organizations and native structures within the cell is required. Such an analysis needs the observation of macromolecules \textit{in situ}, which has been recently made possible with cryo-Electron Tomography (cryo-ET). Cellular cryo-ET is an imaging technology that helps to visualize 3D structures of subcellular structures from a series of 2D projections generated through an electron microscope in cryogenic temperature, where the native structure and orientation of sub-cellular components in the cell are preserved (\cite{luvcic2013cryo}).

	\par With the advent of cryo-ET, many 3D structures in the form of tomograms can be swiftly generated for the analysis of macromolecules they contain. The difficulty, however, lies in decoding the structural information of the macromolecular complexes from these tomograms. These reconstructed 3D tomograms have complexities in the form of very low Signal-to-Nose Ratio (SNR) and missing wedge effects, making the structural recovery of these macromolecules inherently difficult. Classification of macromolecules in tomograms in such cases goes a long way towards their structural recovery from tomogram level data. To classify macromolecules, in a tomogram, we first extract subtomograms, sub-volumes of tomograms where a single macromolecule is most likely to be present. These subtomograms are then classified based on the macromolecule they contain. CNN based methods of supervised and semi-supervised classification of subtomograms that excel in accuracy and speed were proposed by \cite{xu2017deep},\cite{che2018improved}, \cite{gao2020dilated} and \cite{liu2019semi}. The classified subtomograms can be then aligned and averaged (\cite{briggs2013structural}) to get a higher resolution average of the corresponding structure.

	While obtaining large-scale experimental data in the form of tomograms for classification is not an issue, the annotation of this data with corresponding macromolecule identifiers requires a considerable amount of time and compute. Common methods of annotation like template matching proposed by \cite{best2007localization}, \cite{beck2009visual} and  \cite{kunz2015m} are extremely time-consuming and require quality control in the form of inspection by experts, effectively making the entire process long-drawn and laborious. An apparent lack of scalability in data processing makes large-scale data of macromolecular complexes in the form of labeled subtomograms hard to obtain. To get around this issue, methods like unsupervised template-free subtomogram classification have been proposed by \cite{xu2012high}, \cite{bartesaghi2008classification}, and \cite{chen2014autofocused} while automatic annotation procedures using neural networks have been proposed by \cite{chen2017convolutional}. Active learning-based methods that make use of minimal labeled data have also been proposed by \cite{du2021active}

	\par A potential solution to the scarcity of annotated data would be the development of classification algorithms that can classify subtomograms assisted by domain adaptation and randomization algorithms.  Domain randomization methods work by training a classification algorithm on labeled simulated data that is easily available and randomizing this data so that the network trained on it can regularise easily to real data as proposed by \cite{che2019domain}. Contrary to changing (randomizing) the simulated data, domain adaptation-based methods train the deep learning-based classifier in such a way that it can classify both the simulated and the real data corresponding to the source and target domains. Domain in this context refers to data distributions that are related but different. The difference in data distributions is typically projected in the form of a domain shift in the extracted feature space. The concept of domain shift has been illustrated in Fig \ref {fig:domain}. One of the first adversarial domain adaptation methods for the classification of subtomograms was proposed by \cite{lin2019adversarial} while a few shot method for adaptation was proposed by \cite{yu2021few}. All of these methods of classification depend largely on the data simulation procedure used as it forms the base network trained on the source dataset. While some cryo-ET data simulation methods simulate isolated macromolecules, methods proposed by \cite{liu2020efficient} and \cite{pei2016simulating} can dynamically simulate complete tomograms by packing multiple macromolecular complexes with additional factors in the form of simulated noise, contrast transfer function, and missing wedge effects. 
	
	\begin{figure*}[!t]
    	\centering
    	
    	\subfloat[][]{
    	\includegraphics[width=0.45\columnwidth]{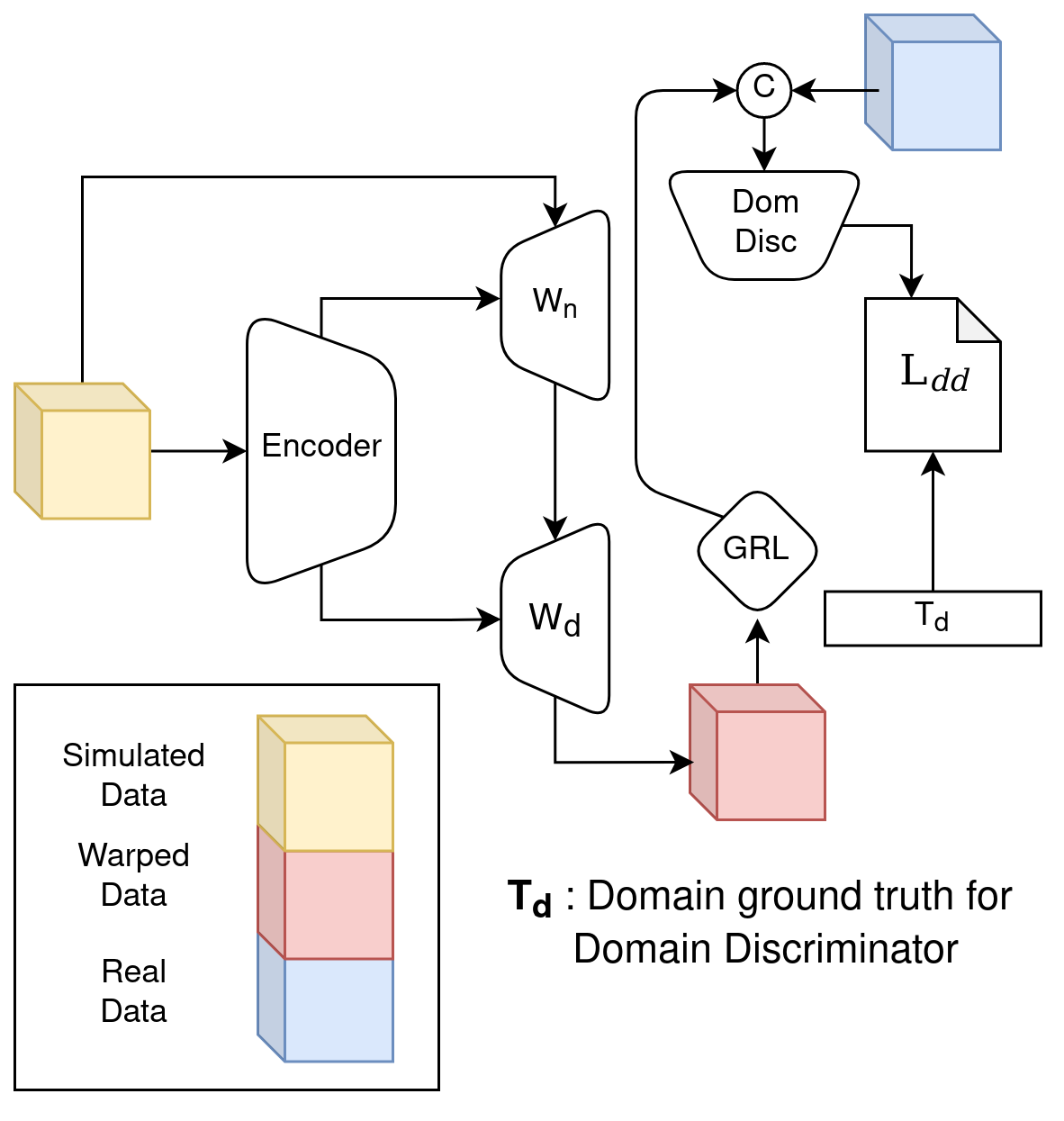}
    	}\hskip 3em
    	\subfloat[][]{
    	\includegraphics[width=0.45\columnwidth]{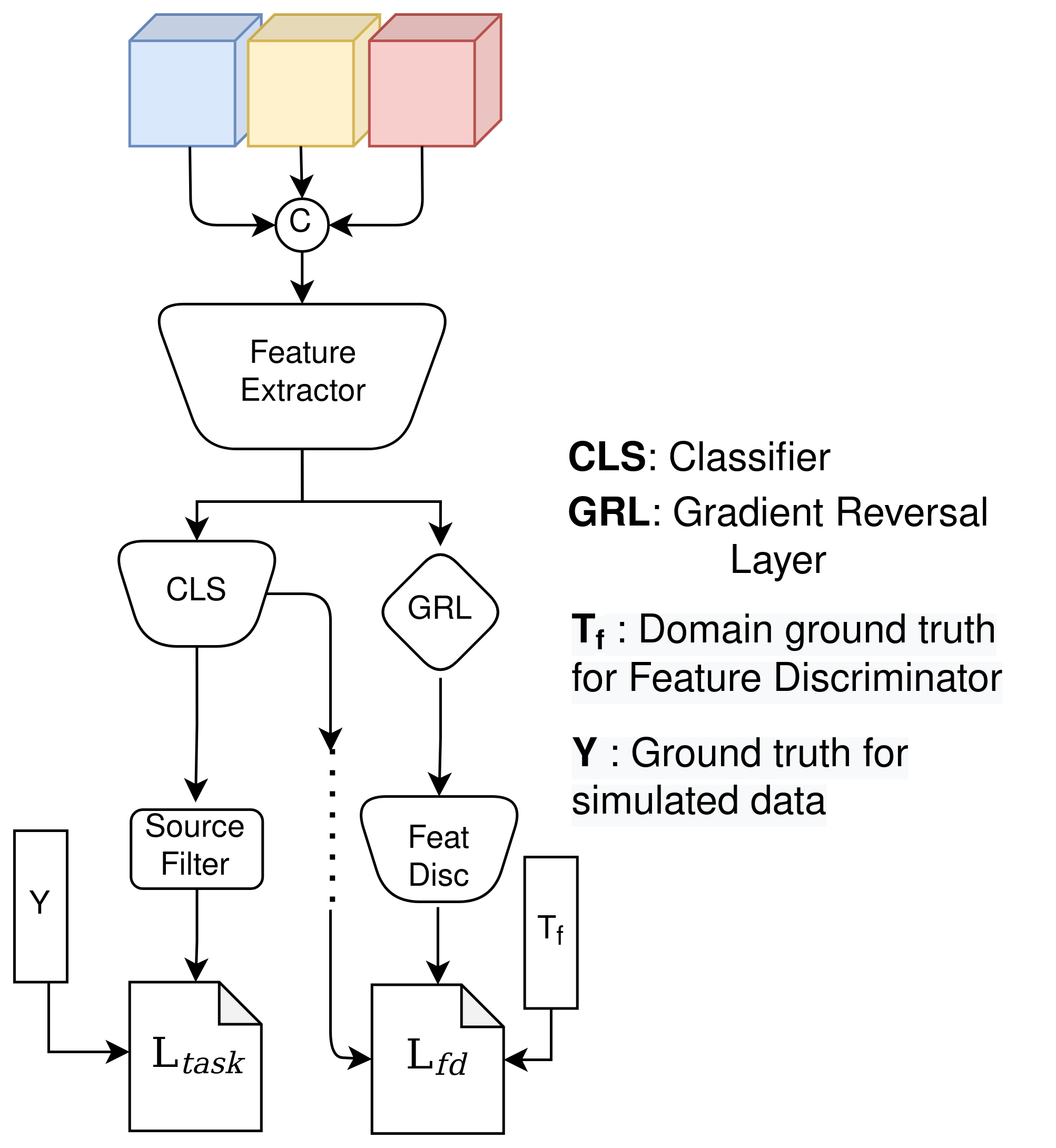}
    	}
    	
    	\label{fig:models}
    	\caption{Model Overview: (a)Diagrammatic Representation of Warp Modules and Domain Discriminator. (b)Diagrammatic Representation of the multi-adversarial domain adaptation framework used. The Source Filter here refers to a filter that allows the classifier to back-propagate on loss calculated only on the simulated and the warped (source) data.}
    \end{figure*}

	In this work, we propose the use of a network-driven domain randomization procedure in conjunction with an unsupervised domain adaptation algorithm for the classification of subtomograms. Fundamentally differing from (\cite{che2019domain}) which focuses on randomizing hyper-parameters and then performing simulations, our network-driven method Cryo-Shift of domain randomization involves gathering information from the already simulated data to add noise and distortions, allowing the model to generalize to the target domain. Furthermore, we propose the use of a multi-adversarial unsupervised domain adaptation framework inspired from \cite{pei2018multi} that takes in minimal unannotated data from the target domain to bridge the domain shift between the simulated and real datasets. We follow the same data generation procedure as \cite{che2019domain} in our experiments.

	\par Our primary contributions, thus, can be summed up as :
	
	\begin{itemize}
		\item The development of a network-driven domain randomization algorithm inspired by \cite{zakharov2019deceptionnet} along with ``warp" modules that distort and alter the simulated data for helping the model generalize better.
		\item The addition of an adversarial loss function with a discriminator to help the warp network provide ``realistic" warps.
		\item The addition of an unsupervised domain adaptation algorithm that helps reduce the domain shift between real and simulated subtomograms, improving cross-domain classification. Being completely unsupervised, the algorithm provides an efficient solution to the problems faced in cryo-ET subtomogram annotation.
		
	\end{itemize}
	
		\section{Methodology}
		\subsection{Overview and Notation}
		Our methods primarily serve to reduce the domain gap between simulated and experimental data via network-guided domain randomization and unsupervised domain adaptation. Our overall pipeline consists of a warp network $W$, a ``step zero network" $M$, a feature discriminator $F_d$, and a domain discriminator $D$. Our data simulation module $S$ takes in a set of hyper-parameters $h$ to simulate data from corresponding density maps ($d_{\textrm{map}}$) of $n$ classes (in our analysis $n=4$). As the first step of our pipeline, we train the base network $M$, consisting of a feature extractor $F_e$ and a classifier $C$, with simulated data $S^{(h)}(d_{\textrm{map}})$ for predicting class labels $\hat{y}$. The feature extractor $F_e$ denotes the convolutional blocks in $M$ while the classifier $C$ refers to the linear layers after $F_e$ in $M$.
	
	    As the second step of the pipeline, we train the warp network $W$, the feature discriminator $F_d$, the domain discriminator $D$ and retrain the base network $M$ simultaneously. The warp network $W$ is fed with simulated data $S^{(h)}(d_{\textrm{map}})$ that it distorts to return a warped set. The domain discriminator $D$ and the warp network $W$ are trained adversarially where $D$ tries to classify the outputs of $W$ as simulated and $W$ tries to fool $D$ into making wrong decisions. The feature extractor from the base network $M$, $F_e$, is fed with a concatenated dataset consisting of the output of the warp network $W(S^{(h)}(d_{\textrm{map}}))$, the simulated data $S^{(h)}(d_{\textrm{map}})$ and a subset $d_{r1}$ of the real data $D_{R}$. While the labels of the simulated data and the output from the warp network are automatically present, the labels of $d_{r1}$ are not used here. Additionally, the output from the warp network is detached from the computational graph before being fed to $F_e$. The output from $F_e$ is fed to the feature discriminator $F_{d}$ and the classifier $C$. Here $F_{d}$ and $F_e$ are trained adversarially while $C$ is trained only on the labeled subset of the data from the feature extractor $F_e$. While $F_{d}$ tries to classify the features from $F_e$ into real and simulated domains, $C$ tries to classify the features into classes accurately.  An important difference between the domain discriminator $D$ and the feature discriminator $F_{d}$ is that the former works directly on subtomograms while the latter works on features. Thus, while the former tries to change the subtomogram to look more like real subtomograms, the latter makes sure that the features learned by the network are similar in nature. The overall algorithm is described in Algorithm \ref{alg:our_algo}. Fig. \ref{fig:models} provides a diagrammatic representation of our method architecture. 
		\begin{algorithm}
			
			\hspace*{\algorithmicindent}\textbf{Input:} hyper-paramters $h$, Density map $d_{\textrm{map}}$, Experimental Data $D_{R}$\\
			\hspace*{\algorithmicindent}\textbf{Output} Trained Network $M$
			\begin{algorithmic}[1]
				\caption{Overall Algorithm}\label{alg:our_algo}
		
				\Procedure{train}{$h$,$d_{\textrm{map}}$,$D_{R}$}
				
				\State Load $M$ \Comment{Pre-trained Step-Zero Network}
				\State $d_{r1},d_{r2}\gets$split($D_{R}$) \Comment{1:9 split} 
				\For{i in num\_iterations}
				\State $W_{\textrm{out}}\gets W(S^{(h)}(d_{\textrm{map}}))$
				\Comment{Domain Randomization}
				\State $d_{s}\gets concat(W_{\textrm{out}},S^{(h)}(d_{\textrm{map}}),d_{r1})$
				\State $(feat,cls\_prob)\gets M(d_{s})$
				\State $F_{\textrm{pred}}\gets F_{d}(feat)$
				\State $D_{\textrm{pred}}\gets D(concat(W_{\textrm{out}},sample(d_{r1})))$

				\State $F_{\textrm{Loss}}\gets \mathcal{L}_{fd}(F_{\textrm{pred}},gt_{f})* cls\_prob$
				\Comment{Adaptation}
				\State $C_{\textrm{Loss}}\gets \mathcal{L}_{\textrm{task}}(cls\_prob[source],gt_{c})$
				\State $D_{\textrm{Loss}}\gets \mathcal{L}_{dd}(D_{\textrm{pred}},gt_{d})$

				\State $W \gets W_{\textrm{step}}$
				\State $M \gets M_{\textrm{step}}$
				\State $D \gets D_{\textrm{step}}$
				\State $F \gets F_{\textrm{step}}$
				
				\EndFor
				\State \textbf{return} $M$\Comment{Trained network}
				\EndProcedure
			\end{algorithmic}
		\end{algorithm}

		\subsection{Step Zero Network}
		
		The step zero network or the base network is a simple classification network we pre-train on the simulated data $S^{(h)}(d_{\textrm{map}})$. This network forms the baseline upon which further training takes place with the help of warp modules and adaptation algorithms. To establish the consistent nature of out algorithm, we make use of multiple architectures in our ablation experiments (in Section \ref{expt}) while proceeding with a single architecture, CB3D from (\cite{che2018improved}), in the main experiment. A list of the model architectures used along with their corresponding parameters is provided in Table \ref{tab:networks}. 
		\begin{table}[!t]
			\caption {Networks used}
			\begin{tabular*}{\textwidth}{c @{\extracolsep{\fill}} c}
				\toprule
				\textbf{Architecture}&\textbf{Parameter Count}\\\midrule
				CB3D (\cite{che2018improved}) &72.2 M\\
				DSRF3D\_v2 (\cite{che2018improved}) &18.2 M\\
				DenseNet3D (\cite{hara3dcnns}) &11.2 M\\
				ResNet3D (\cite{hara3dcnns}) &85.1 M\\

				\bottomrule
			\end{tabular*}
			\label{tab:networks}
			
		\end{table}
		
		\subsection{Warp Network}
		\par The Warp Network $W$ is built of a set of warp modules that distort or alter the simulated data and help the base network generalize more to the experimental data during inference. Domain randomization in the context of the warp modules we use is a misnomer because while domain randomization modules deal with randomizing the source domain to help the model generalize better to any target domain, we specialize our warp network to work on a particular target domain. Although this restricts the portability of our model onto any target dataset, this helps us enhance the performance of our network on the single target dataset we train it on. Thus, while \cite{zakharov2019deceptionnet} proposes making use of a gradient reversal layer (GRL) to oppose the classifier for training the randomization network, we propose the use of a discriminator with a GRL to oppose our warp network. Here, we pass the output of the warp network and randomly sampled data from the validation set in equal proportions to the discriminator that is tasked with classifying the data as real or fake. The presence of the GRL between the warp network and the discriminator makes the warp network inclined to fool the discriminator into classifying the warped data as real. The concept of GRL, inspired from \cite{ganin2014unsupervised} is that the gradients are reversed during backpropagation, thereby allowing the Discriminator and the warp network to train in opposing manners.

		\par The warp network consists of an encoder-decoder based architecture with the output of a single encoder $E$ being fed to multiple decoders, each emulating one warping module. The encoder consists of 3 convolutional blocks with skip connections that connect to the decoder layers and are concatenated there. The input to the encoder is batch-wise 3D cryo-ET subtomogram data in the form of $I_{n}\in\mathbb{R}^{b\times 1 \times 40 \times 40 \times 40}$ with b denoting the batchsize. The output from the encoder, the bottleneck $B_{e}$ can be expressed as $O_{e}\in\mathbb{R}^{b\times 256\times 5 \times 5 \times 5}$. This bottleneck is split into two equal parts channel-wise and fed to the two decoders which act as warp modules.
		The warping modules we make use of are:
		\begin{itemize}
    		\item Noise Module ($W_n$)
			\item Distortion Module ($W_d$)
		\end{itemize}
		
		\subsubsection{Noise Module}
		The noise module ($W_n$) adds noise to the input subtomogram $I_{n}$ as the first randomization step. The encoder split is then passed through a simple decoder with 3 convolutional blocks where the skip connections from the encoder are concatenated at their respective layers. The output from the last decoder block is passed through a scaled sigmoid activation function, restricting its value to $(-0.1,0.1)$. The final output $d_{0}\in\mathbb{R}^{b\times 1 \times 40 \times 40 \times 40}$ is the summed up with the input to the module $I_{n}$ and is represented as $O_{n} \in\mathbb{R}^{b\times 1 \times 40 \times 40 \times 40}$.

		\subsubsection{Distortion Module}
		The distortion module takes in two inputs, $O_n$ from the noise module and a channel-wise split of the encoder output $E_{dec}$. $E_{dec}$ is fed to two subnetworks: the first consisting of a single convolutional and linear layer while the second being a decoder with 3 convolutional blocks with the corresponding skip connections from the encoder concatenated. The first network provides an output in the form of a single value per sample per channel $\alpha \mathbb{R}^{b\times 3}$ restricted to $(0,1)$, while the second decoder gives out a 3D dense grid $G \in \mathbb{R}^{b\times 3 \times 40 \times 40 \times 40}$. This dense-grid is then convolved with a $7 \times 7$ Gaussian 3D kernel having $\sigma=1$ and multiplied in a channel-wise fashion with $\alpha$ where the 3 channels store values of x, y, and z coordinates. The values in this new dense-grid $grid_{d}$ are then mapped with the coordinates in $O_n$ and the corresponding deformed output subtomogram $O_d$ is created.
		\begin{equation}
		grid_{d}=\text{conv}(W_{d1}(E_{dec}),G_{7\times7})*W_{d1}(E_{dec})
		\end{equation}
		\begin{equation}
		O_{d}=\text{map}(O_{n},grid_{d})
		\end{equation}

		\subsection{Domain Discriminator}
		The Domain discriminator $D$ consists of an encoder module, acting as a binary classifier, and is pre-trained on simulated data and the unlabelled $d_{r1}$ subset of real data $D_{R}$. The discriminator predicts if the data it is supplied with is from the real dataset or the simulated dataset. While it strives to reduce the discriminative loss $\mathcal{L}_{dd}$ and classify the domains correctly, the warp network training alongside adversarially, tries to fool it in classifying its output as real.
		
		\subsection{Domain Adaptation}
		
		We propose an unsupervised domain adaptation paradigm inspired by \cite{pei2018multi}. This paradigm consists of $n$ similar feature discriminators where $n$ represents the number of classes our classifier has. These discriminators are simply binary classifiers that try to classify the features to their respective domains (simulated or real). A concatenated set consisting of output from the warp network, simulated subtomograms, and real subtomogram data ($d_{r1}$) in equal proportions is fed to the feature extractor $F_e$ of base model $M$ at each iteration. These features from $F_e$ are fed to each of the $n$ feature discriminators. The domain labels for the simulated and warped data are zero, while the domain label for the experimental data is set to one. A weighted gradient reversal layer (GRL) is set up between the discriminators and the feature extractor, forming an adversarial relationship between the networks during training. During the calculation of the mean discriminative loss for training the feature discriminators, a weighting system is used with the predicted class probabilities from the classifier $C$ (detached from the computational graph), for the same features, acting as weights as demonstrated in equation \ref{eq:lfd}. The pre-trained classifier, which can already predict real subtomograms with considerable accuracy, comes in use now as the discriminative loss for a particular discriminator is multiplied with the probability of the subtomogram belonging to that particular class as predicted by the classifier.

		\subsection{Objective Function}
		The step-zero network, $M$, is trained on the domain randomized simulated data $S^{(h)}$ using $\mathcal{L}_{\textrm{task}}$, a categorical cross-entropy loss as the loss function. We make use of this pre-trained model $M$ in the next stage where we simultaneously train the base network $M$, the domain discriminator $D$, the warp network $W$, and the feature discriminator $F_{d}$. The discriminative loss $\mathcal{L}_{\textrm{disc}}$ is a combination of the domain discriminative loss $\mathcal{L}_{dd}$ used to train $D$ and $W$ and the feature discriminative loss $\mathcal{L}_{fd}$ used to train $F_{d}$ and $F_e$. As in the previous stage, in this stage also a task loss $\mathcal{L}_{\textrm{task}}$ is used to train the step-zero network $M$ ($F_e$ and $C$). The corresponding loss function for the entire algorithm can be expressed as:
		\begin{equation}
		\mathcal{L}_{\textrm{train}}=\mathcal{L}_{\textrm{task}}(M(W(S^{(h)}(d_{\textrm{map}}))),y)+\mathcal{L}_{\textrm{disc}}
		\end{equation}
		\begin{equation}
		\mathcal{L}_{\textrm{disc}}=\mathcal{L}_{dd}(D,t_{d})+\mathcal{L}_{fd}(F,t_{f})
		\end{equation}
		where $y$ represents the task ground truth. $t_{d}$ and $t_{f}$ represent the domain ground truth for the two discriminators.

		The components of $\mathcal{L}_{\textrm{disc}}$, $\mathcal{L}_{fd}$ and $\mathcal{L}_{dd}$  can be represented as:
		\begin{equation}
		\mathcal{L}_{fd}=\sum_{i=1}^{N}\sum_{j=1}^{X}
		(t_{fj}log(F_{i}(y_{j}))-(1-t_{fj})log(1-F_{i}(y_{j})))prob_{ji}
		\label{eq:lfd}
		\end{equation}
		\begin{equation}
		\mathcal{L}_{dd}=\textrm{BCELoss}(D,t_{d})
		\end{equation}
		where $F_{i}$ represents the individual discriminators from 1 to $N$ with $N$ denoting the total classes (four in our case). $y_{j}$ represents the one dimensional features from $F_{e}$ for the $j^{th}$ sample with $X$ denoting the total samples in a batch. $prob_{j}$ represents the class confidence of the $i^{th}$ class for the $j^{th}$ sample.

	\begin{table}[!htbp]
	
		\caption{Ablation Studies in Domain Randomization on CB3D}
		\begin{tabular*}{\textwidth}{c @{\extracolsep{\fill}} cc}
			\toprule
			Modules Used & Loss Function & Macro Avg.\\
			\midrule
			$W_{n}+W_{d}$& $\mathcal{L}_{task}^{*}$ &0.72\\
			$W_{n}$& $\mathcal{L}_{task}+\mathcal{L}_{dd}$ &0.72\\
			$W_{d}$& $\mathcal{L}_{task}+\mathcal{L}_{dd}$ &0.75 \\
		
			\midrule
			$\mathbf{W_{d}+W_{n}}$& $\mathbf{\mathcal{L}_{task}+\mathcal{L}_{dd}}$ &\textbf{0.77} \\
			\bottomrule
		\end{tabular*}
		\footnotesize{$^*$The warp network was trained on reverse grad with the classifier}
		\label{tab:abl_expt_1}
	\end{table}
	
	\subsection{Training}\label{expt}
	We follow the methodology described above and pre-train the step-zero network as a basic classification module. We use Adam as an optimizer with a learning rate of 1e-4 and set betas to 0.9 and 0.999. We make use of a similar optimizer in the second phase of the training with the learning rate set to 1e-5 for the step zero network and the learning rates for the warp network, domain discriminator, and the feature discriminator set to 5e-3, 1e-6, and 1e-5 respectively. Amongst these, the optimizer of the base network is monitored with its learning rate reduced when the loss plateaus. The loss checked for this plateau condition is the task loss $\mathcal{L}_{\textrm{task}}$ on the validation set we constructed earlier and used as an unannotated data source. We train the step zero network for 120 epochs in the first stage and the 50 epochs in the next stage of the pipeline.

\begin{table}[!t]

	\caption{Classwise accuracy comparison with step-zero networks}
	\begin{tabular*}{\textwidth}{c @{\extracolsep{\fill}} ccccc}
		\toprule
		Architecture&\textbf{$C_{0}$}&\textbf{$C_{1}$}&\textbf{$C_{2}$}&\textbf{$C_{3}$}&Avg.\\\midrule
		CB3D &0.70 &0.63 &0.61 &0.91 &0.71\\
		\textbf{CB3D (W)} &\textbf{0.71} &\textbf{0.75} &\textbf{0.69} &\textbf{0.92} &\textbf{0.77}\\
		
		DSRF3D\_v2 &0.69 &0.64 &0.59 &0.91 &0.71\\
		
		\textbf{DSRF3D\_v2 (W)} &\textbf{0.73} &\textbf{0.66} &\textbf{0.67} &\textbf{0.89} &\textbf{0.74}\\

		DenseNet3D &0.70 &0.54 &0.53 &0.97 &0.69\\
		
		\textbf{DenseNet3D (W)} &\textbf{0.84} &\textbf{0.73} &\textbf{0.45} &\textbf{0.96} &\textbf{0.74}\\
		
		ResNet3D &0.77 &0.50 &0.54 &0.91 &0.68\\
		
		\textbf{ResNet3D (W)} &\textbf{0.70} &\textbf{0.57} &\textbf{0.59} &\textbf{0.96} &\textbf{0.70}\\
		\bottomrule
    \label{tab:abl_expt_2}
	\end{tabular*}
\end{table}
	
\section{Experiments}
\subsection{Simulated Dataset}
The simulated data we use in our work is generated using a data generation process similar to (\cite{che2019domain}). The data generation process starts with density maps obtained directly from EMDB (Electron Microscopy Data Base) accessions or running situs package (\cite{wriggers1999situs}) on RCSB PDB (Protein Data Bank) files.  In our experiments, density maps of four distinct classes (Ribosome, Proteosome, TRiC, and Membrane) were obtained from EMDB accessions. These maps have a shape of $40\times40\times40$ with a voxel spacing of 1.368nm. To simulate the missing wedge effect, the 3D structural data is projected on a series of 2D projection images with varying tilt angles based on the Maximum Wedge Angle hyperparameter specified. This projection data is then convolved with Contrast Transfer Function (CTF) and Modulation Transfer Function (MTF), adjusted by hyperparameters like Defocus and spherical aberration, for the simulation of the respective optical effects. At this stage, Gaussian Noise is added to the data to simulate the corresponding SNR level and the data is back-projected to obtain the 3D subtomogram. This 3D data is randomized by sampling hyperparameters uniformly from the following list, while in the specific case of the SNR, the sampling is done at a logarithmic scale.

\begin{figure}
			\centering
			\subfloat[][Ribosome]{\includegraphics[width=0.2\columnwidth]{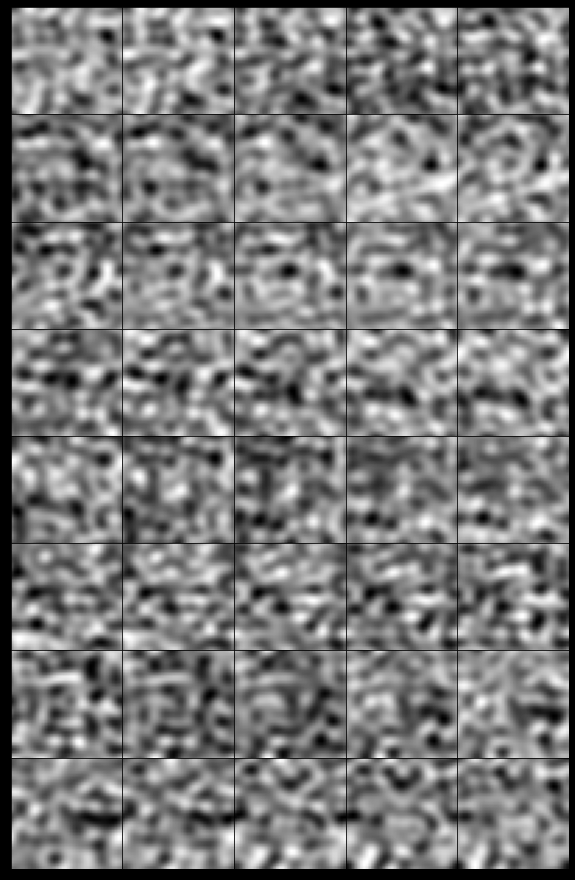}}
			\subfloat[][Proteasome]{\includegraphics[width=0.2\columnwidth]{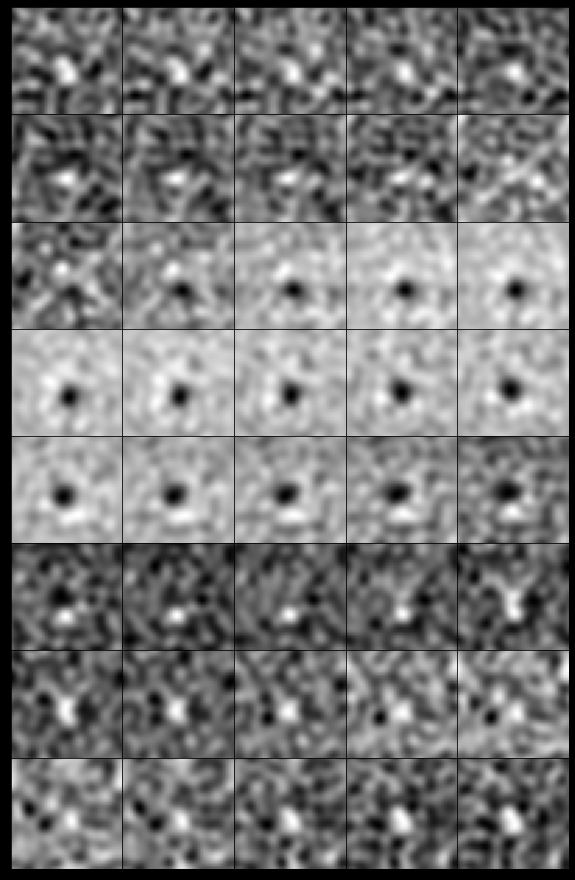}}
			\subfloat[][TRiC]{\includegraphics[width=0.2\columnwidth]{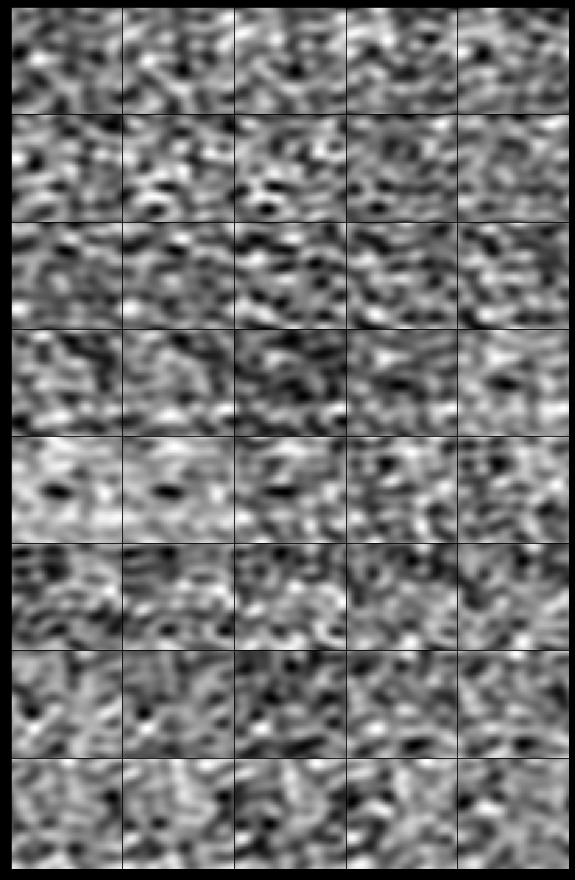}}
			\subfloat[][Membrane]{\includegraphics[width=0.2\columnwidth]{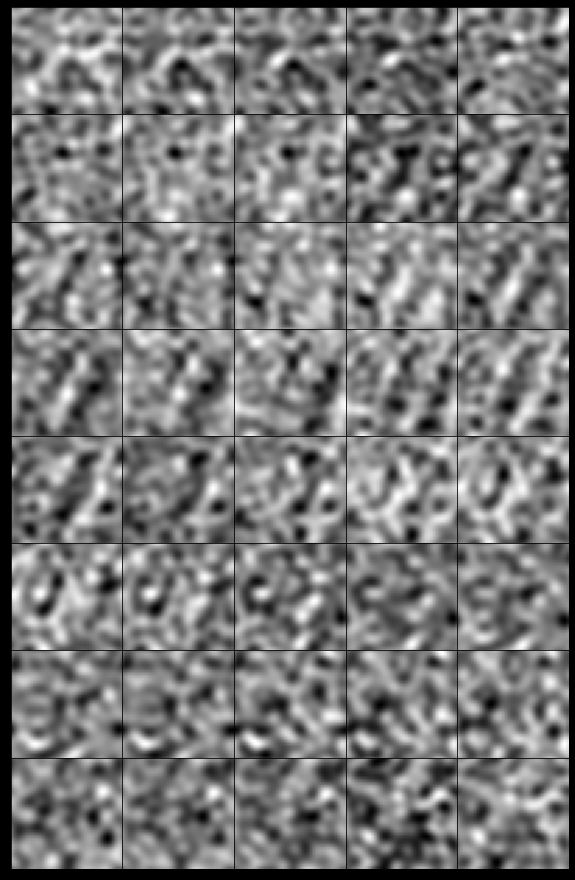}}
			\caption{\label{fig:demo_sim} 2D slice visualization of simulated subtomograms}
		\end{figure}

\begin{itemize}
	\item SNR: 0.03 to 10
	\item MWA: 0\degree to 50\degree
	\item Dz: -12 to 0
	\item Cs: 1.5 to 3.0
\end{itemize}

\noindent where MWA stands for Maximum Wedge Angle and Cs and Dz are hyperparameters for Spherical Abberation and Defocus respectively. A sample from the simulated data is provided in Fig. \ref{fig:demo_sim}. Further samples demonstrating the qualitative effect of these hyperparameters on the simulation is available in the supplementary.

\subsection{Experimental dataset}
\begin{figure}[!h]
			\centering
			\subfloat[][Ribosome]{\includegraphics[width=0.2\columnwidth]{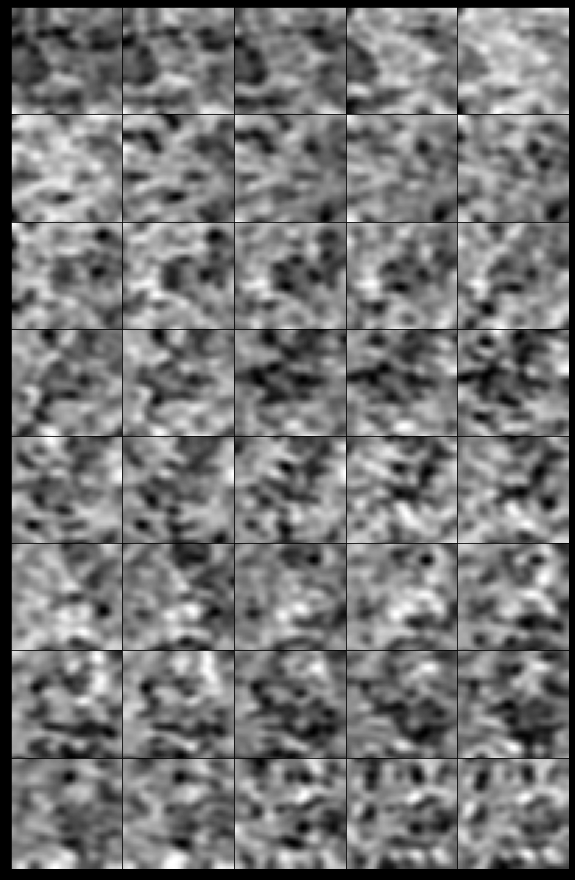}}
			\subfloat[][Proteasome]{\includegraphics[width=0.2\columnwidth]{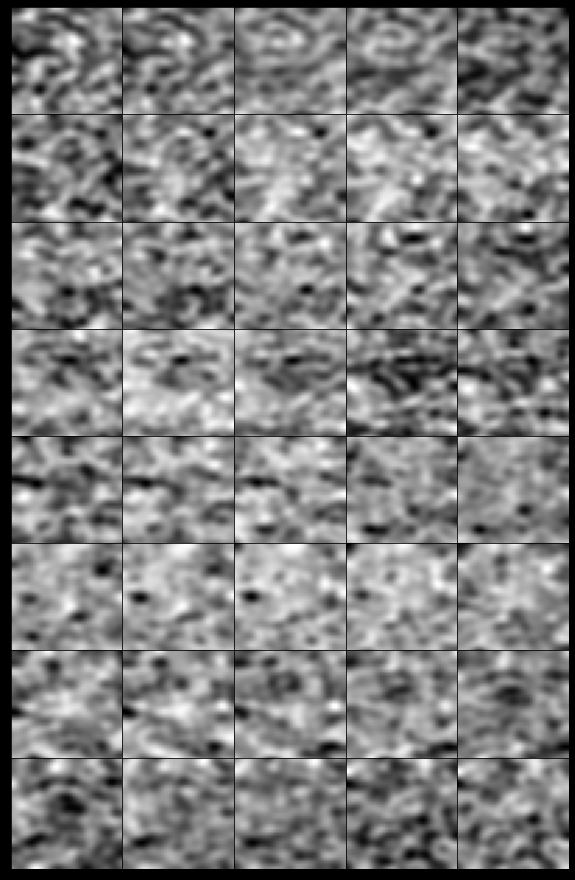}}
			\subfloat[][TRiC]{\includegraphics[width=0.2\columnwidth]{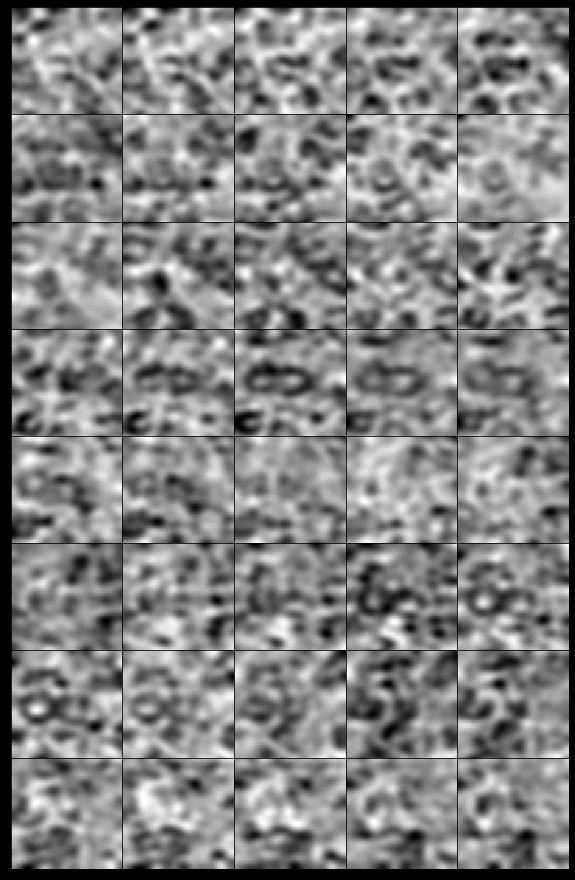}}
			\subfloat[][Membrane]{\includegraphics[width=0.2\columnwidth]{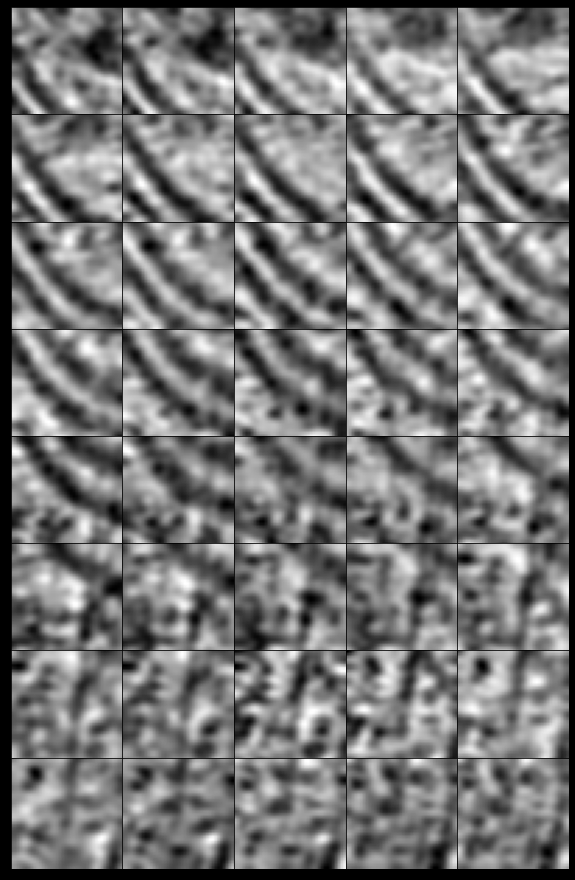}}
			\caption{\label{fig:demo} 2D slice visualization of experimental subtomograms}
		\end{figure}

The real dataset we use in our experiments consists of 1051 subtomograms, which have been processed by template search and have been filtered manually from rat neuron tomograms (\cite{guo2018situ}). The data has a tilt angle range of -50\degree to +70\degree and is divided into 4 classes, namely ribosome, mitochondrial membrane, TRiC, and single capped proteasome. A summary of the class-wise distribution of the data in terms of the number of subtomograms is available in Table \ref{tab:data}. A figure containing slices of a subtomogram belonging to each class is available in Fig. \ref{fig:demo}. This data is further split in a 1:9 ratio where the smaller part of the split is fed to the network as unlabelled real data for randomization and adaptation. Since this smaller subset of the data is passed as unlabelled and does not actively participate in reducing the classification loss, we reuse it as a validation set in our experiments. 

\begin{table}[!t]
			\caption{Experimental data}
			\begin{tabular*}{\textwidth}{c @{\extracolsep{\fill}} ccc}
				\toprule
				\textbf{Class Label}&\textbf{Subtomograms}&\textbf{Validation Set}&\textbf{Test set}\\\midrule
				Ribosome ($C_{0}$)&80 &8 &72\\
				Proteasome ($C_{1}$)&386 &38 &348\\
				TRiC ($C_{2}$)&125 &12 &113\\
				Membrane ($C_{3}$)&460 &46 &414\\
				\midrule
				Total &1051 &104 &947\\
				\bottomrule
			\end{tabular*}
			\label{tab:data}
		\end{table}
		
\subsection{Ablation Study}
We perform ablation experiments for the warp network, as the feature discriminator part of our algorithm does not involve multiple modules. For our ablation studies, we use multiple neural networks varying in depth to demonstrate the efficacy of our methods on more than one network architecture. The network architectures we use along with their trainable parameter counts are listed in \ref{tab:networks}. Table \ref{tab:abl_expt_2} presents the performance of the warp module along with the performance of corresponding step zero networks (in terms of macro-average accuracy across 4 classes). The presence of ``(W)" along with the neural network architecture used denotes that the model has been trained with the warp network, while the absence of this notation implies the performance recorded is that of the pre-trained step zero network. An increasing average accuracy while using the warp modules is expected here as we are not only using the domain randomization algorithm to fool the discriminator, but are also training the classifier to keep up with it. Table \ref{tab:abl_expt_1} presents more ablation studies using CB3D as the base network with modules zeroed out at places to demonstrate their individual contributions to the network. Table \ref{tab:abl_expt_1} also includes a study where we remove the domain discriminator and force the warp modules to train against the classification loss as proposed in \cite{zakharov2019deceptionnet}. Our ablation studies demonstrate the efficacy of each module in the warp network along with individual components of the loss function we use. Additional ablation experiments on simulated data only are provided in the supplementary.

\begin{figure}
			\centering
			\subfloat[][Input]{\includegraphics[width=0.2\columnwidth]{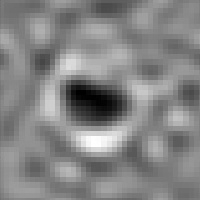}}
			\subfloat[][Output from $W_{n}$]{\includegraphics[width=0.2\columnwidth]{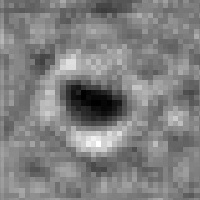}}
			\subfloat[][Output from $W_{d}$]{\includegraphics[width=0.2\columnwidth]{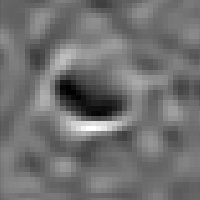}}
			\caption{\label{fig:warp_out} Individual affects of the warp modules on a subtomogram slice}
\end{figure}

\section{Results}		
\begin{figure}[!h]
	\centering
	\subfloat[][t-SNE Before Adaptation]{\includegraphics[width=0.5\columnwidth]{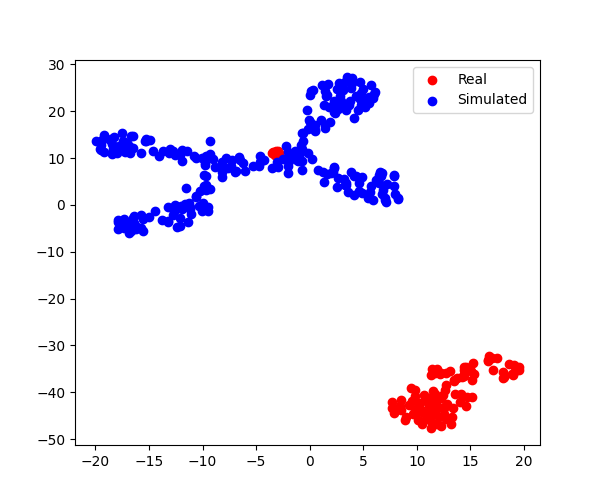}}
	\subfloat[][t-SNE After Adaptation]{\includegraphics[width=0.5\columnwidth]{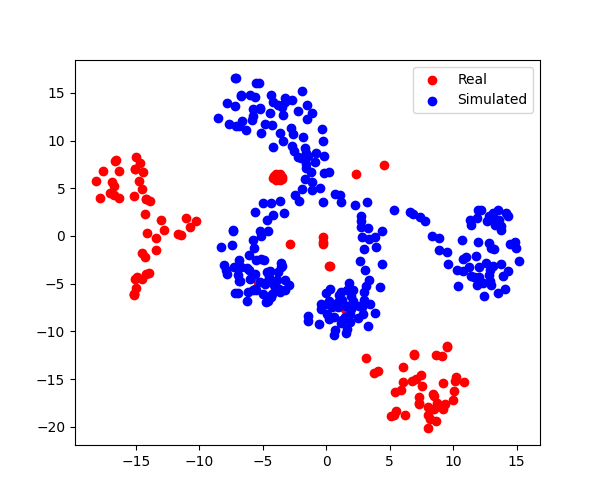}}
	\label{fig:tsne}
	\caption{Unsupervised Domain Adaptation Results}
\end{figure}

The extensive results of randomization and adaptation on the CB3D model are presented in Table \ref{tab:final_cb3d}. We compare our results with relevant domain adaptation methods like FSADA (\cite{yu2021few}) and FADA (\cite{motiian2017fewshot}) in Table \ref{tab:comp}. While these are domain adaptation methods that make use of labels in limited amounts, we propose a completely unsupervised procedure of the same and outperform these methods by a large margin (7 \%). To further ensure a fair comparison with these methods, we train all of them on the same architecture as proposed by \cite{yu2021few}. We also include t-SNE plots in Fig \ref{fig:tsne} that represent feature spaces extracted from $M$ to visualize the reduction of domain shift between simulated and experimental data brought about by our training methodology. The t-SNE plots give a visual of how a domain shift is present between the features of the real and the simulated data. It further demonstrates how a reduction in the same has been achieved in the features extracted from our network. 

\begin{table}[!t]
\caption{Adaptation Results on CB3D}
\begin{tabular*}{\textwidth}{c @{\extracolsep{\fill}} cccc}
\toprule
\textbf{Label}        & \textbf{Precision} & \textbf{Recall} & \textbf{F1\_score} & \textbf{Support} \\ \midrule
{Ribosome}     & {0.65}      & {0.99}   & {0.78}      & {72}      \\
{Proteasome}   & {0.90}       & {0.88}   & {0.89}      & {348}     \\
{TRiC}         & {0.83}      & {0.78}   & {0.80}       & {113}     \\
{Membrane}     & {0.99}      & {0.94}   & {0.96}      & {414}     \\
\midrule
{Accuracy}     & {}          & {}       & {0.90}       & {947}     \\
\textbf{Macro Avg}    & \textbf{0.84}      & \textbf{0.90}    & \textbf{0.86}      & \textbf{947}     \\
{Weighted Avg} & {0.91}        & {0.90}    & {0.90}       & {947}     \\ \bottomrule
\end{tabular*}
\label{tab:final_cb3d}

\end{table}

\begin{table}[!t]
	
	\caption{Accuracy comparison with other Domain Adaptation Methods}
	\begin{tabular*}{\textwidth}{c @{\extracolsep{\fill}}| ccccc}
		\toprule
		Methods&\textbf{$C_{0}$}&\textbf{$C_{1}$}&\textbf{$C_{2}$}&\textbf{$C_{3}$}&Avg.\\\midrule

		FADA 3-shot (\cite{motiian2017fewshot}) &0.62 &0.70 &0.66 &0.91 &0.72\\
		FADA 5-shot &0.61 &0.76 &0.64 &0.91 &0.73\\
		FADA 7-shot &0.78 &0.64 &0.62 &0.91 &0.73\\
		FSDA 3-shot (\cite{yu2021few}) &0.61 &0.70&0.67&0.87&0.71\\
		FSDA 5-shot &0.64&0.78&0.66&0.92&0.75\\
		FSDA 7-shot &0.65&0.78&0.70&0.94&0.77\\
		\midrule

		\textbf{Cryo-shift$^*$} &\textbf{0.75} &\textbf{0.85} &\textbf{0.83} &\textbf{0.95} &\textbf{0.84}\\
		\bottomrule
	\end{tabular*}
	\footnotesize{*Trained with our methodology on architecture from FSDA}
	\label{tab:comp}
\end{table}

\section{Conclusion}
The study of \textit{in situ} macromolecular complexes enabled with cryo-electron tomography is essential for the analysis of various cellular processes. Shortcomings in the imaging procedure in conjunction with high structural complexity in tomograms make this analysis difficult to perform. The classification of subtomograms as sub-volumes of these tomograms forms an integral step in the structural recovery of macromolecules along with aiding in their analysis. Supervised classification methods, however, require large-scale annotated data that is extremely scarce. In this work, we propose an unsupervised domain adaptation and randomization framework which can help train neural networks completely on simulated data and unlabeled experimental data. We first train a base network on simulated data and then use a network-driven domain randomization framework to help the network make better predictions on the experimental dataset. Additionally, we make use of a multi-adversarial domain adaptation module where we help the classifier generate similar features for both real and simulated data, thus mitigating the domain shift between the real and the simulated data features and helping the model classify better in both of these domains. Our work can be further used for automated annotation of subtomograms after being trained on similar simulated data, thus forming an important step in the completely autonomous recognition and structural recovery of subtomograms with the help of neural networks for better analysis of macromolecules and their interactions.

We find our work can be improved upon with future works aimed at further development of simulation algorithms to bridge the domain gap between real and simulated data. Unsupervised deep learning based algorithms for the analysis of genomic data (\cite{liu2021simultaneous}) have made considerable progress with the help of generative networks. A possibility lies in the use of novel conditional generative networks (\cite{vahdat2021score,liu2021density}) to generate class-specific cryo-et data for improving upon the algorithms demonstrated in this work. 

\section*{Acknowledgment}

We thank Dr. Qiang Guo for sharing with us experimental rat-neuron tomograms that we make use of as real data.

\section*{Funding}

This work was supported in part by U.S. National Institutes of Health (NIH) grants R01GM134020 and P41GM103712, U.S. National Science Foundation (NSF) grants DBI-1949629 and IIS-2007595, AMD COVID-19 HPC Fund, and the Mark Foundation For Cancer Research 19-044-ASP. X.Z. was supported in part by a fellowship from Center of Machine Learning and Health at Carnegie Mellon University.

\bibliography{ref}

\pagebreak
\begin{center}
	\textbf{\large Supplemental Material: Cryo-Shift}
\end{center}
\setcounter{equation}{0}
\setcounter{figure}{0}
\setcounter{table}{0}
\setcounter{page}{1}
\makeatletter
\renewcommand{\theequation}{S\arabic{equation}}
\renewcommand{\thefigure}{S\arabic{figure}}
\renewcommand{\bibnumfmt}[1]{[S#1]}
\renewcommand{\thetable}{S\arabic{table}}


An example of the simulation hyperparameters and their effect on the data being simulated is demonstrated in Figures S1-S8. To provide a qualitative analysis, we show results at the extremes of a particular hyperparameter range, while maintaining the other hyperparameters at the mid-points of their ranges. Since SNR is sampled at a logarithmic scale, we take the midpoint of the logarithmic scale for SNR in these demonstrations. In these diagrams, $C_0$, $C_1$, $C_2$, $C_3$ refer to the four classes Ribosome, Proteasome, TRiC, and Membrane respectively. We further provide some experiments on purely simulated data to gauge the performance of our model against the baseline step zero network (CB3D). In these experiments detailed in Table \ref{tab:exp}, we simulate the data at specified SNR levels and try to adapt our network to predict classes at a different SNR level with the help of network driven domain randomization and adaptation. The top row in each cell of the table represents the performance of the baseline network while the bottom row demonstrates our model's performance.

\begin{table}[!htbp]
	\centering
	\caption{Results on Simulated data}
	\label{tab:exp}
	\begin{tabular}{|c|c|c|c|c|}
		\hline
		& \multicolumn{4}{c|}{Target Domain}\\\hline
		\multirow{6}{*}{Source Domain} & SNR & 0.03 & 0.5 & 10\\\cline{2-5}
		& \multirow{2}{*}{0.03} & \multirow{2}{*}{-} & 0.83 & 0.86\\
		&  &  & 0.79 & 0.84\\\cline{2-5}
		
		& \multirow{2}{*}{0.5} & 0.26 & \multirow{2}{*}{-} & 0.54\\
		&  & 0.28 & & 0.67\\\cline{2-5}
		
		& \multirow{2}{*}{10} & 0.26 & 0.49 & \multirow{2}{*}{-}\\
		&  & 0.28 & 0.76 & \\\cline{2-5}
		
		\hline
		
	\end{tabular}
	
\end{table}

\begin{figure}[!htbp]
	\centering
	\subfloat[][$C_{0}$]{%
		\label{fig:ex3-a}%
		\includegraphics[height=1.6in]{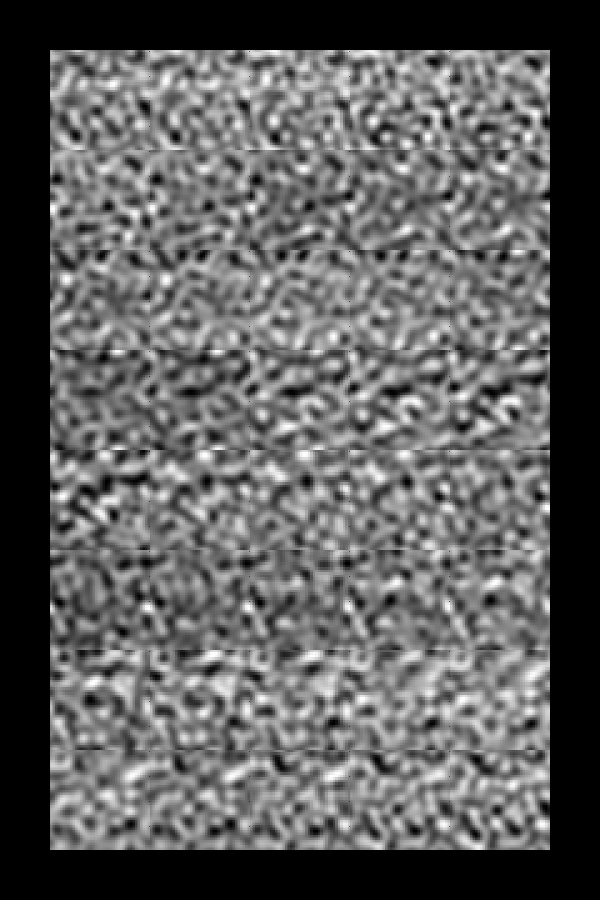}}%
	\hspace{2pt}%
	\subfloat[][$C_{1}$]{%
		\label{fig:ex3-b}%
		\includegraphics[height=1.6in]{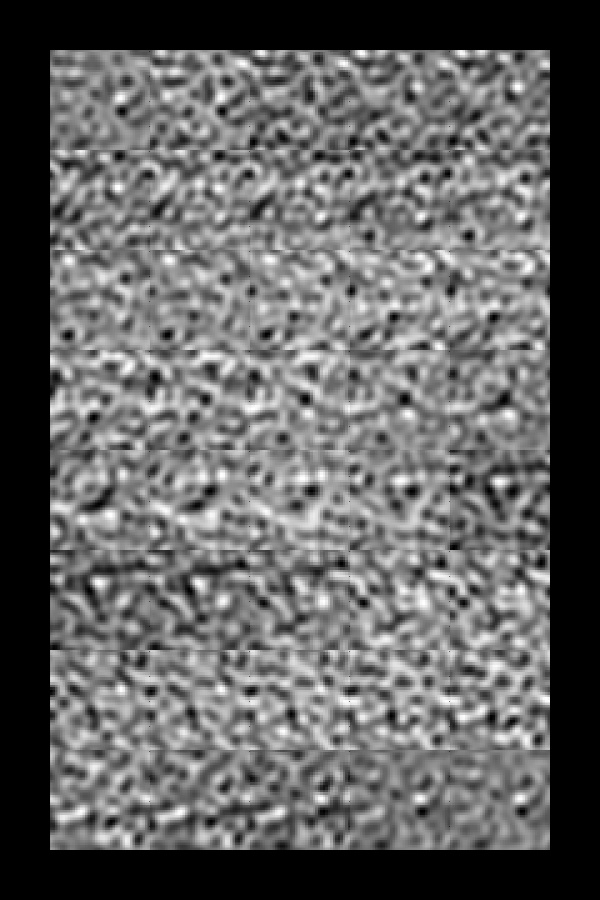}}
	\subfloat[][$C_{2}$]{%
		\label{fig:ex3-c}%
		\includegraphics[height=1.6in]{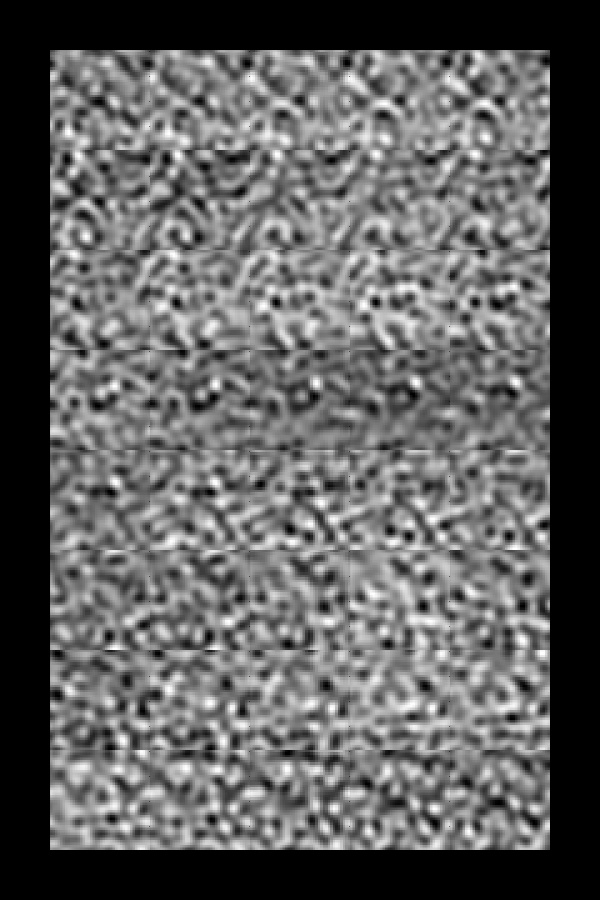}}%
	\hspace{2pt}%
	\subfloat[][$C_{3}$]{%
		\label{fig:ex3-d}%
		\includegraphics[height=1.6in]{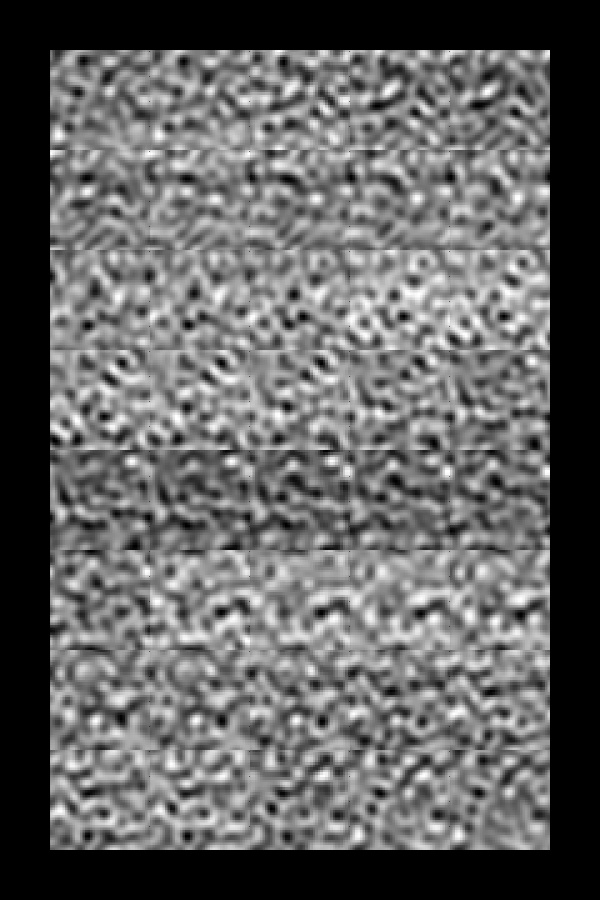}}%

	\caption[]{SNR: 0.03, MWA: 25, Dz: -6, Cs: 2.25 (Simulation Hyperparameters)}
\end{figure}

\begin{figure}[!htbp]
	\centering
	\subfloat[][$C_{0}$]{%
		\label{fig:ex3-a}%
		\includegraphics[height=1.6in]{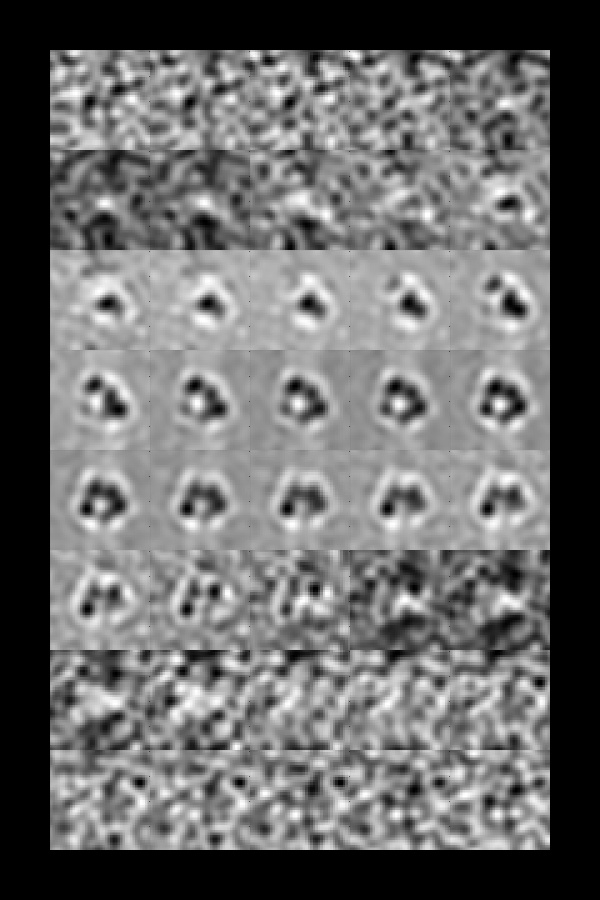}}%
	\hspace{2pt}%
	\subfloat[][$C_{1}$]{%
		\label{fig:ex3-b}%
		\includegraphics[height=1.6in]{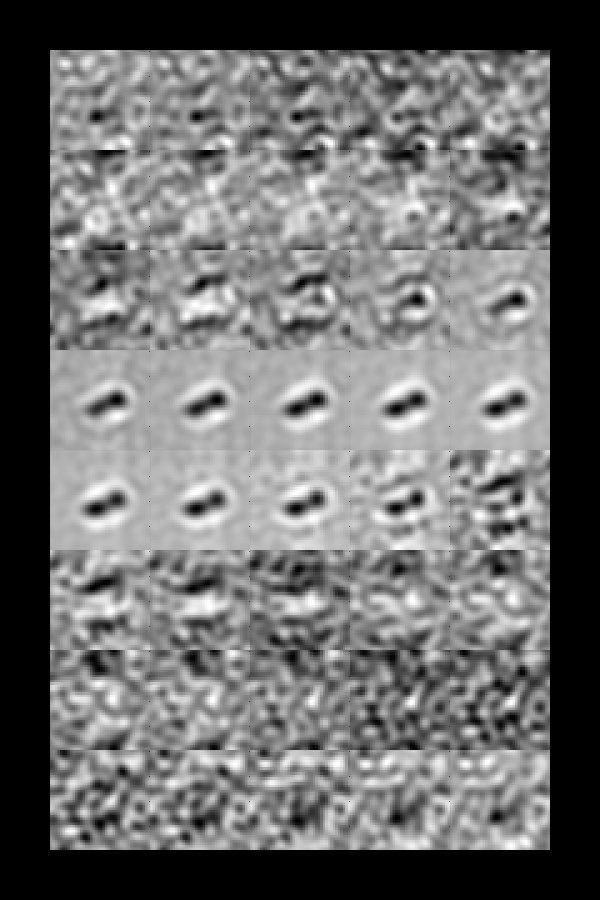}}
	\subfloat[][$C_{2}$]{%
		\label{fig:ex3-c}%
		\includegraphics[height=1.6in]{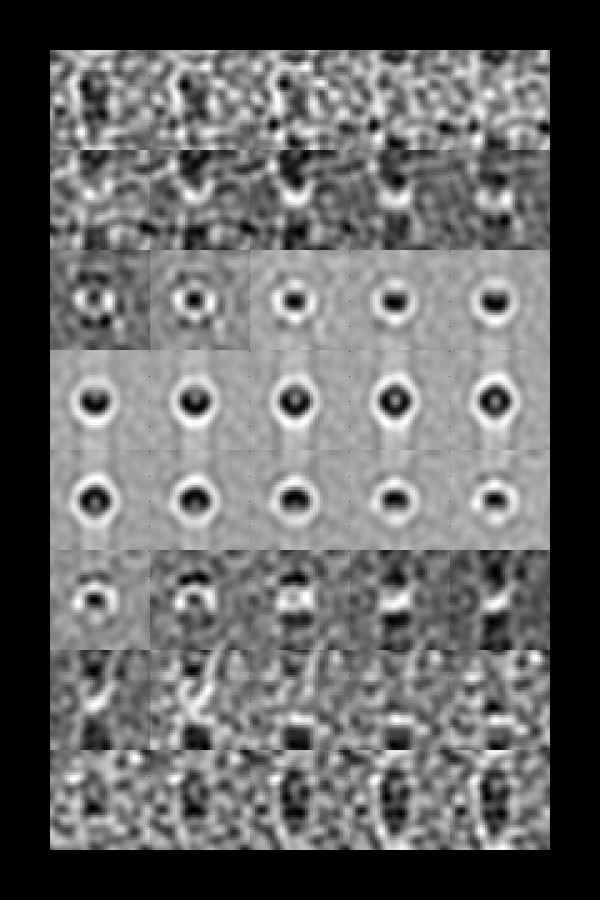}}%
	\hspace{2pt}%
	\subfloat[][$C_{3}$]{%
		\label{fig:ex3-d}%
		\includegraphics[height=1.6in]{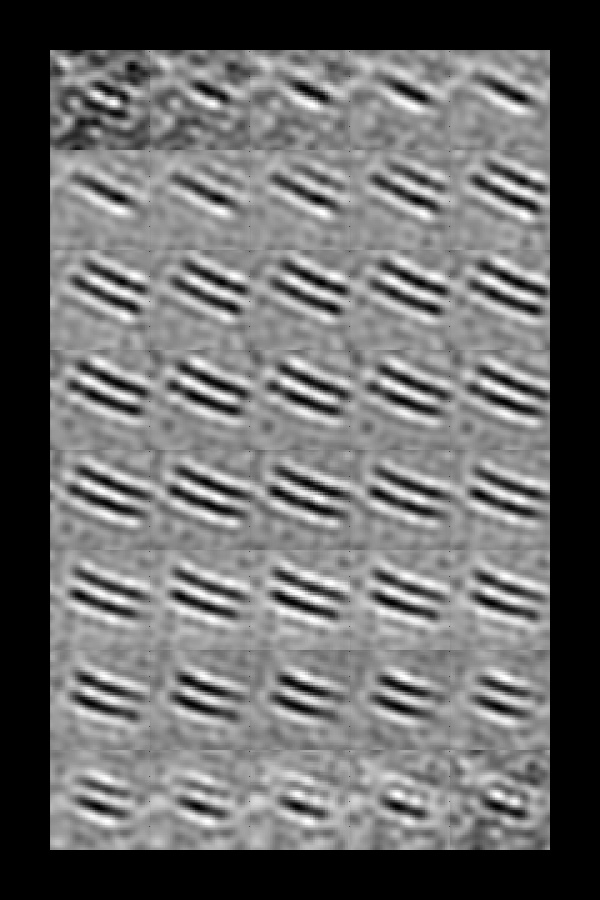}}%

	\caption[]{SNR: 10, MWA: 25, Dz: -6, Cs: 2.25 (Simulation Hyperparameters)}
\end{figure}

\begin{figure}[!htbp]
	\centering
	\subfloat[][$C_{0}$]{%
		\label{fig:ex3-a}%
		\includegraphics[height=1.6in]{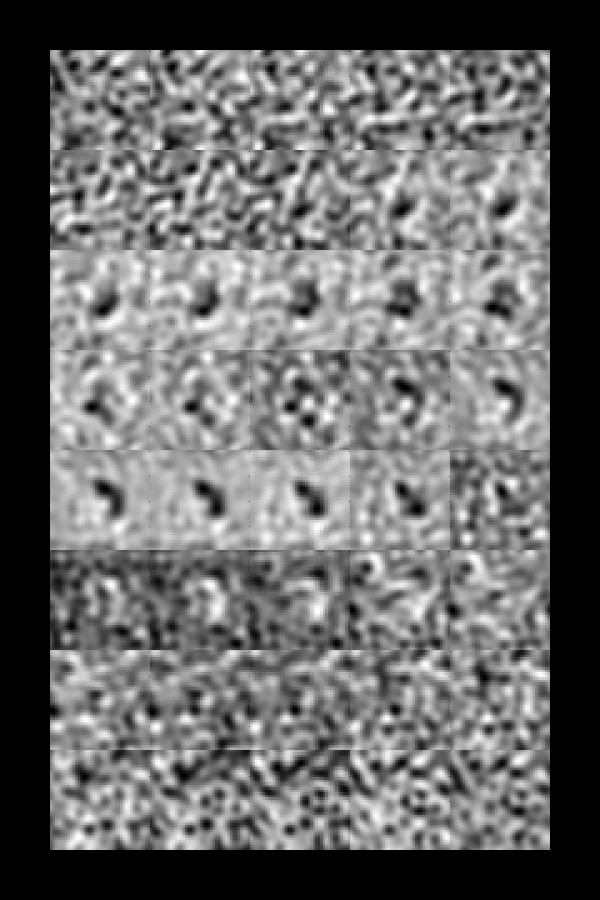}}%
	\hspace{2pt}%
	\subfloat[][$C_{1}$]{%
		\label{fig:ex3-b}%
		\includegraphics[height=1.6in]{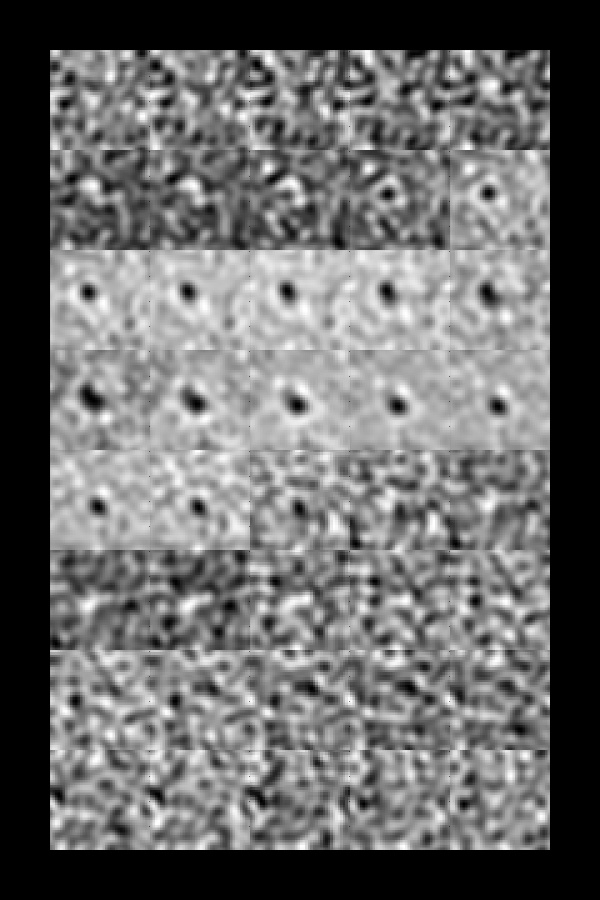}}
	\subfloat[][$C_{2}$]{%
		\label{fig:ex3-c}%
		\includegraphics[height=1.6in]{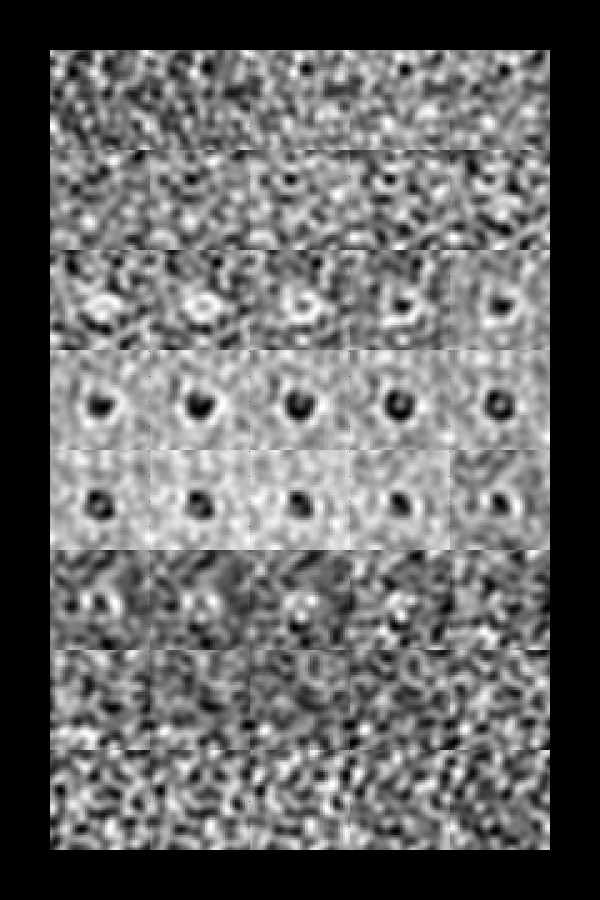}}%
	\hspace{2pt}%
	\subfloat[][$C_{3}$]{%
		\label{fig:ex3-d}%
		\includegraphics[height=1.6in]{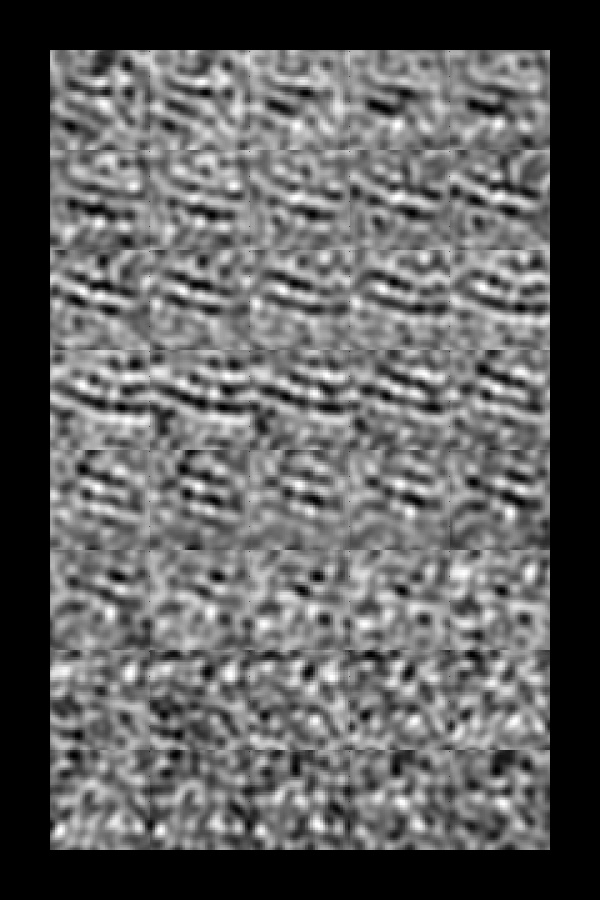}}%

	\caption[]{SNR: 0.547, MWA: 0, Dz: -6, Cs: 2.25 (Simulation Hyperparameters)}
\end{figure}

\begin{figure}[!htbp]
	\centering
	\subfloat[][$C_{0}$]{%
		\label{fig:ex3-a}%
		\includegraphics[height=1.6in]{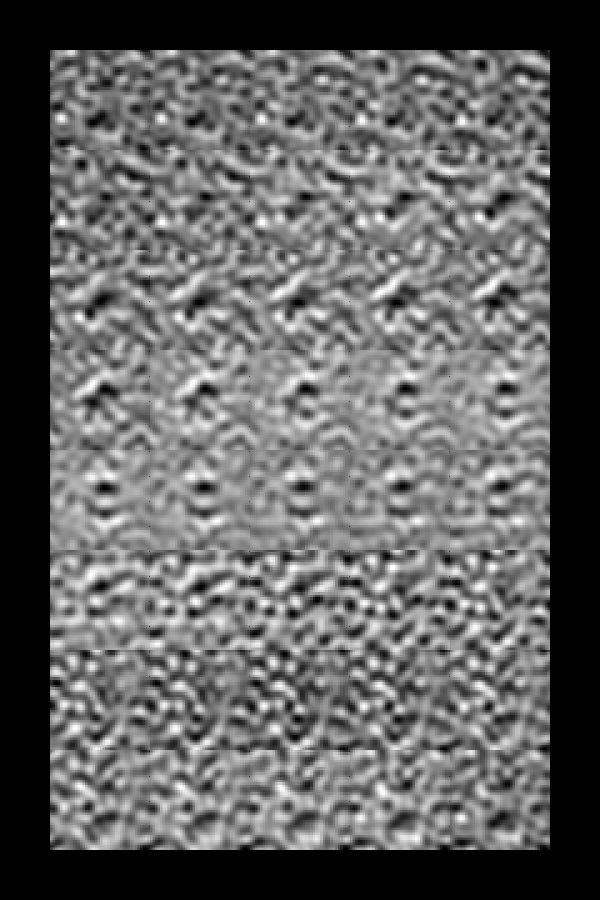}}%
	\hspace{2pt}%
	\subfloat[][$C_{1}$]{%
		\label{fig:ex3-b}%
		\includegraphics[height=1.6in]{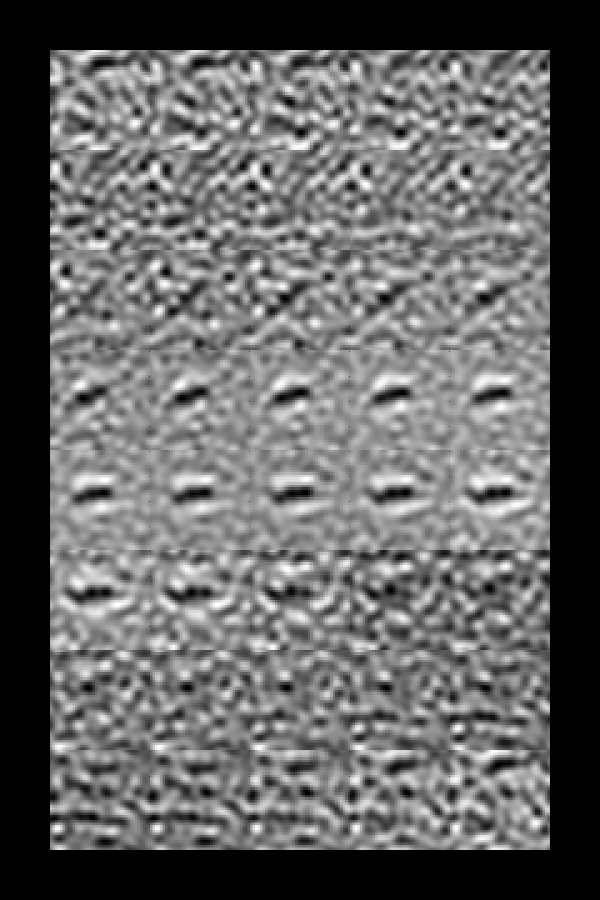}}
	\subfloat[][$C_{2}$]{%
		\label{fig:ex3-c}%
		\includegraphics[height=1.6in]{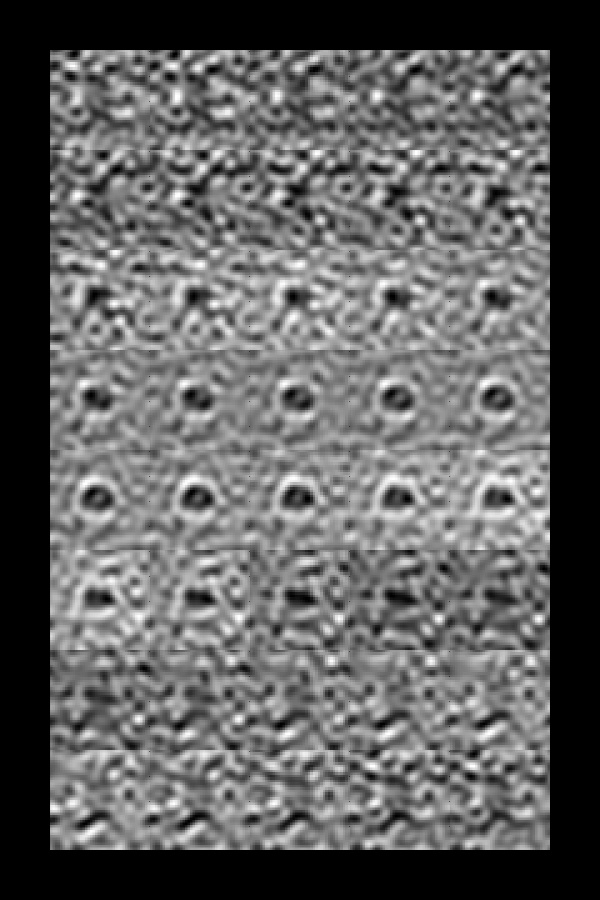}}%
	\hspace{2pt}%
	\subfloat[][$C_{3}$]{%
		\label{fig:ex3-d}%
		\includegraphics[height=1.6in]{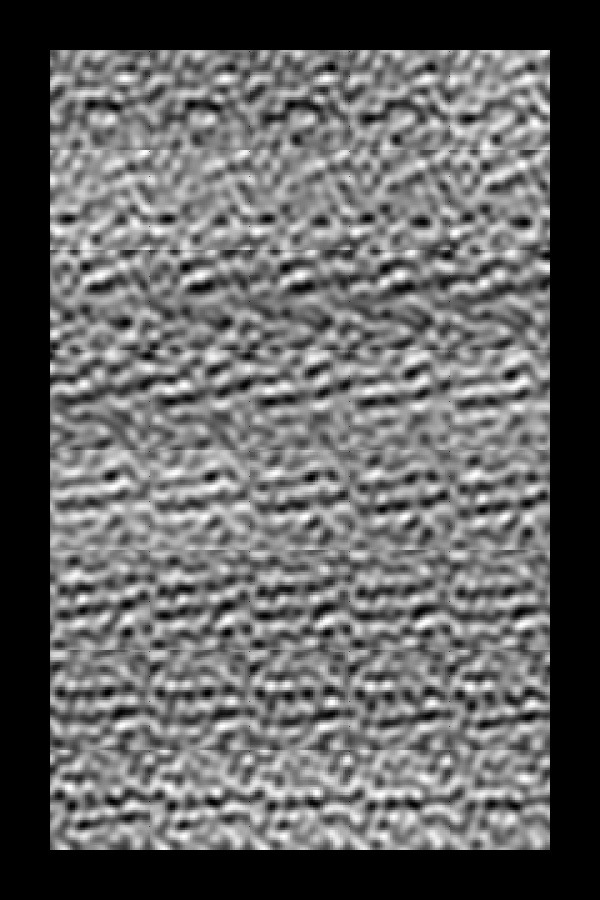}}%

	\caption[]{SNR: 0.547, MWA: 50, Dz: -6, Cs: 2.25 (Simulation Hyperparameters)}
\end{figure}

\begin{figure}[!htbp]
	\centering
	\subfloat[][$C_{0}$]{%
		\label{fig:ex3-a}%
		\includegraphics[height=1.6in]{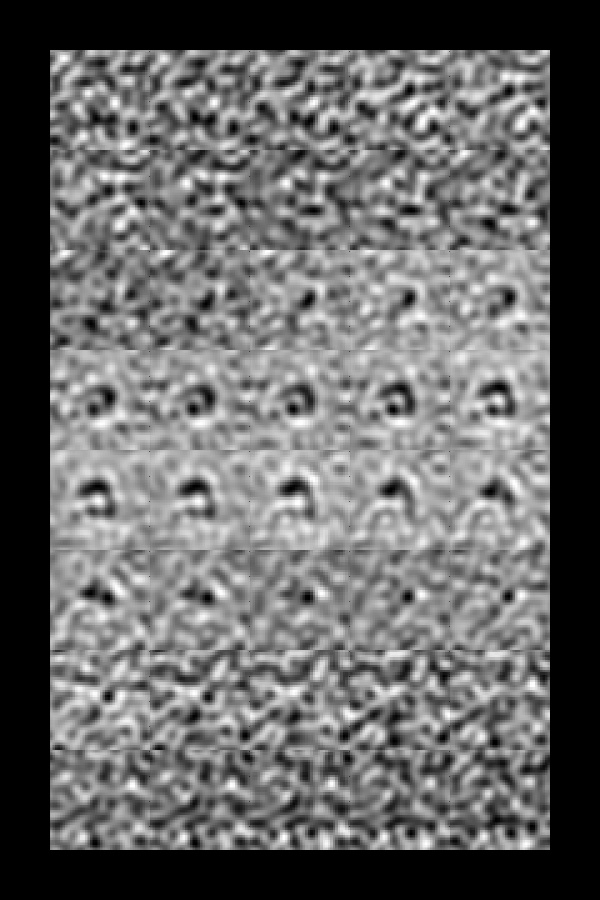}}%
	\hspace{2pt}%
	\subfloat[][$C_{1}$]{%
		\label{fig:ex3-b}%
		\includegraphics[height=1.6in]{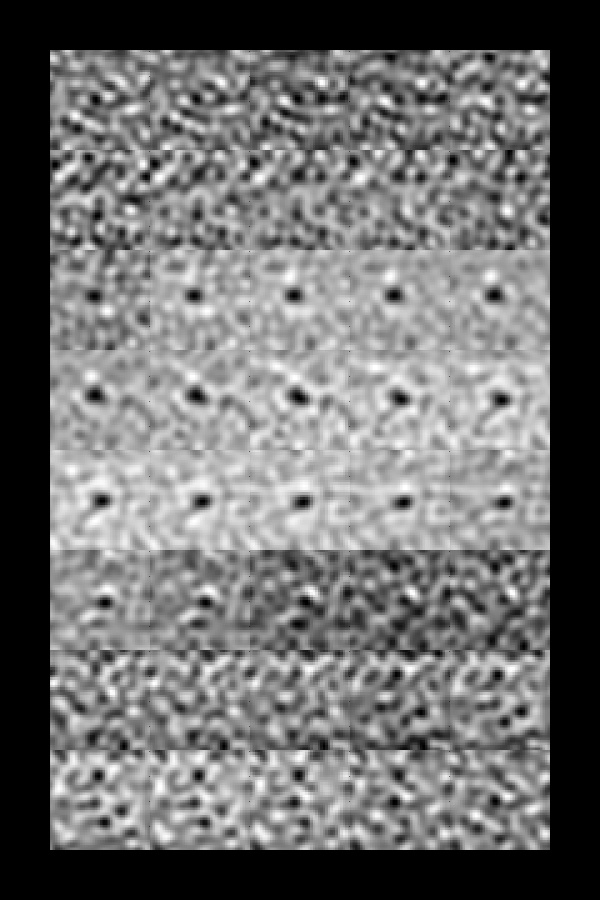}}
	\subfloat[][$C_{2}$]{%
		\label{fig:ex3-c}%
		\includegraphics[height=1.6in]{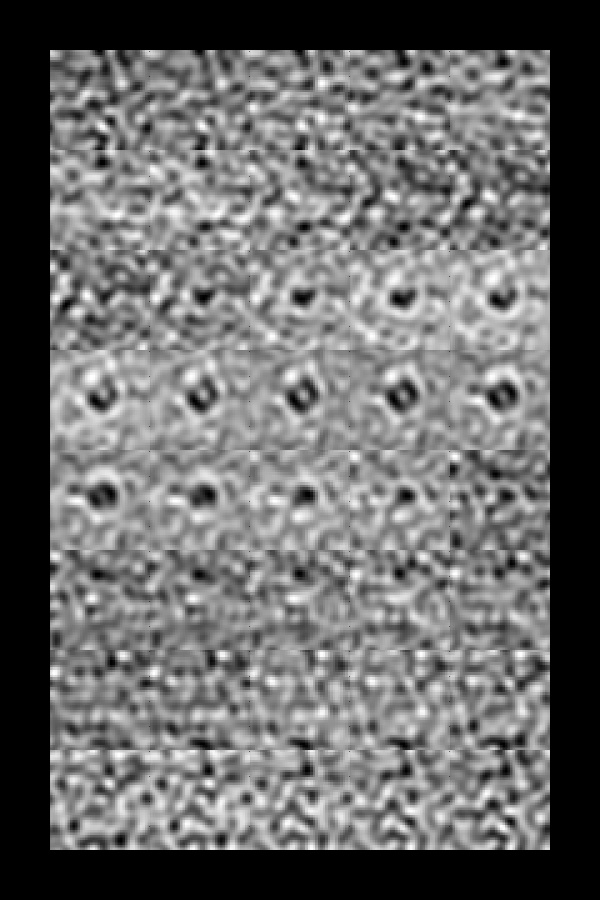}}%
	\hspace{2pt}%
	\subfloat[][$C_{3}$]{%
		\label{fig:ex3-d}%
		\includegraphics[height=1.6in]{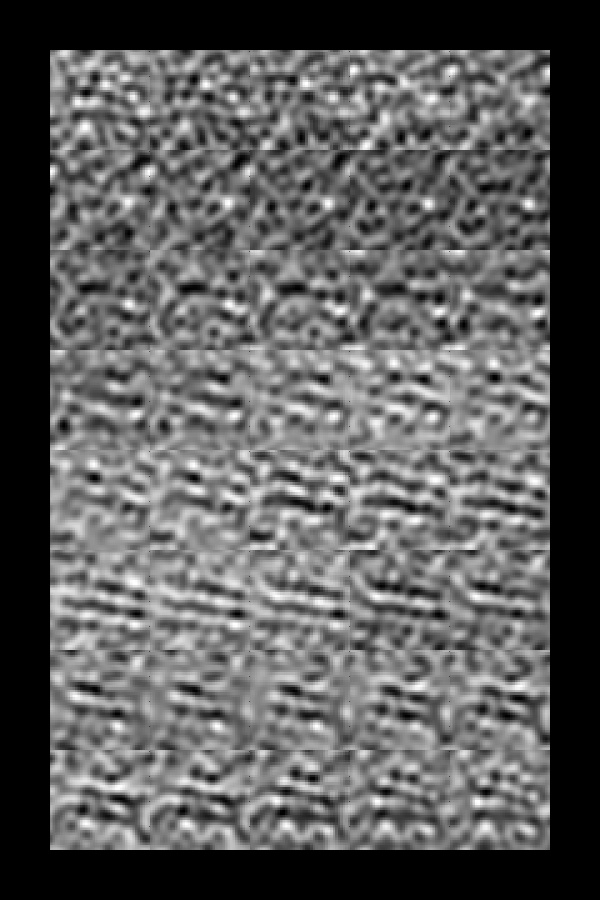}}%

	\caption[]{SNR: 0.547, MWA: 25, Dz: -6, Cs: 1.5 (Simulation Hyperparameters)}
\end{figure}

\begin{figure}[!htbp]
	\centering
	\subfloat[][$C_{0}$]{%
		\label{fig:ex3-a}%
		\includegraphics[height=1.6in]{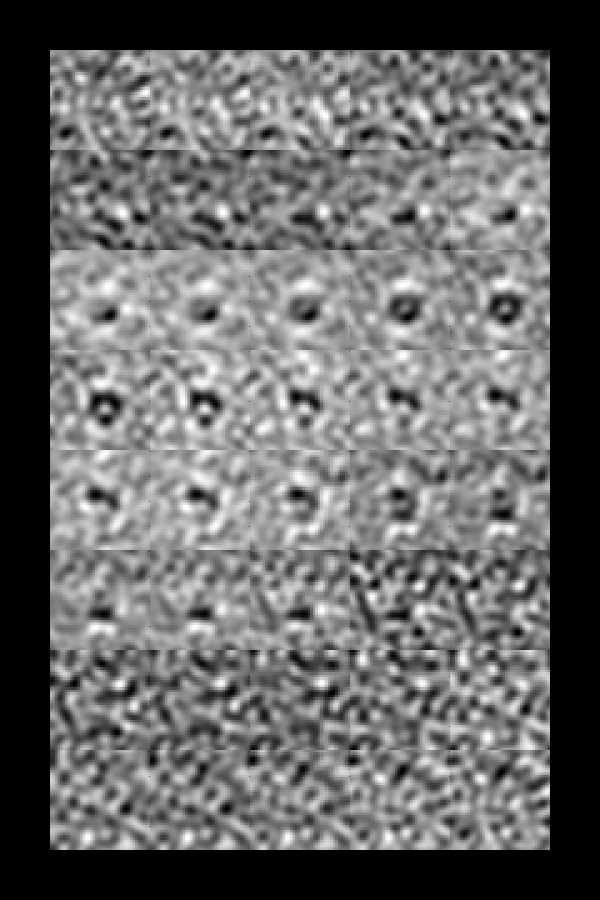}}%
	\hspace{2pt}%
	\subfloat[][$C_{1}$]{%
		\label{fig:ex3-b}%
		\includegraphics[height=1.6in]{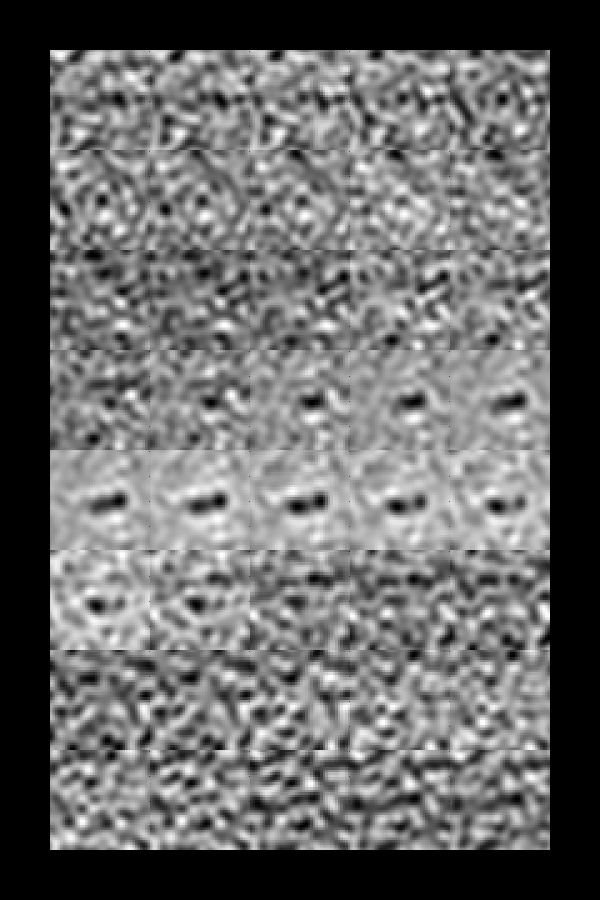}}
	\subfloat[][$C_{2}$]{%
		\label{fig:ex3-c}%
		\includegraphics[height=1.6in]{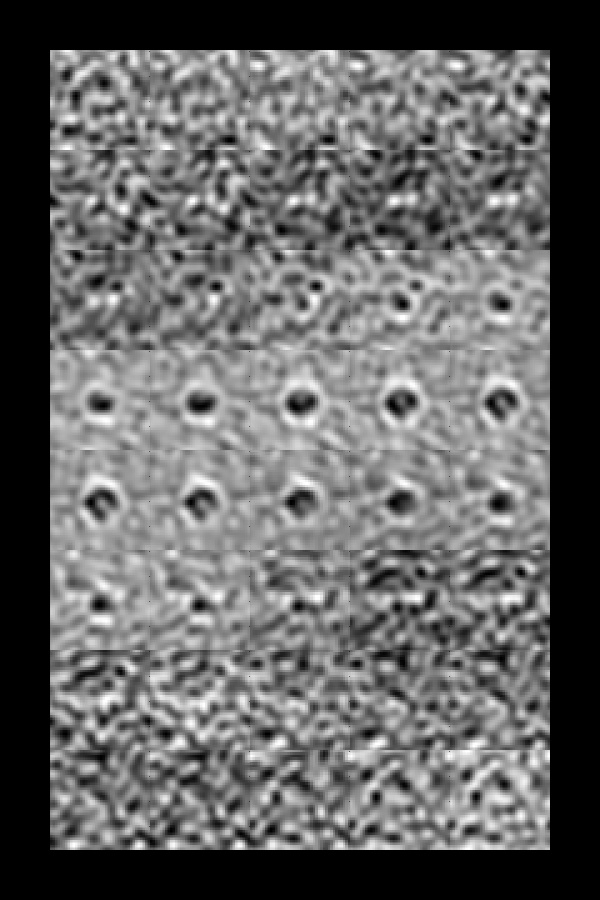}}%
	\hspace{2pt}%
	\subfloat[][$C_{3}$]{%
		\label{fig:ex3-d}%
		\includegraphics[height=1.6in]{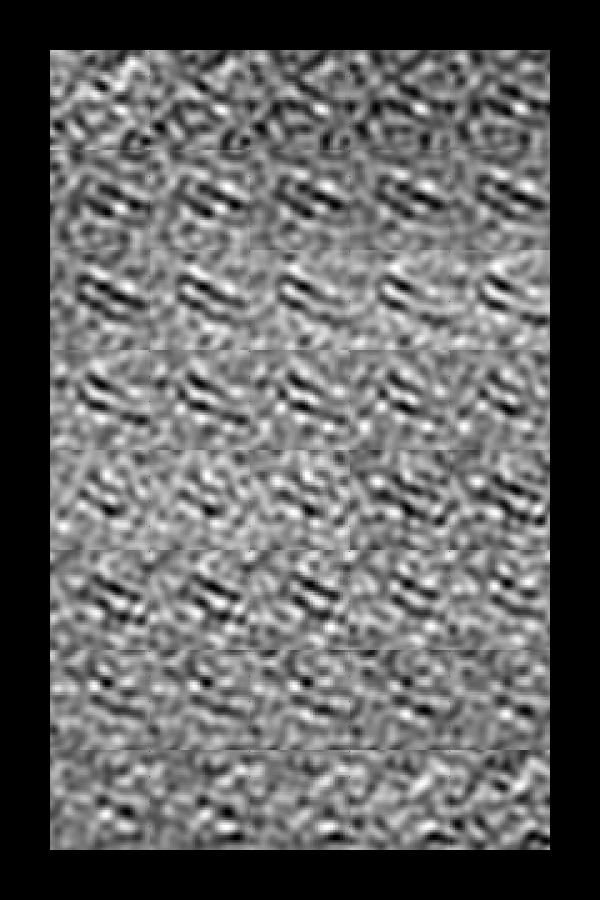}}%

	\caption[]{SNR: 0.547, MWA: 25, Dz: -6, Cs: 3 (Simulation Hyperparameters)}
\end{figure}

\begin{figure}
	\centering
	\subfloat[][$C_{0}$]{%
		\label{fig:ex3-a}%
		\includegraphics[height=1.6in]{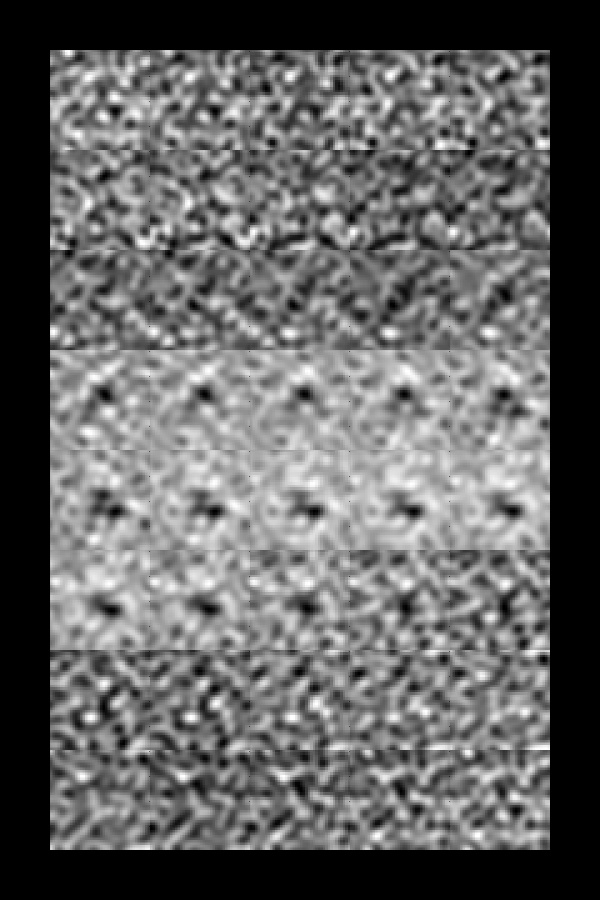}}%
	\hspace{2pt}%
	\subfloat[][$C_{1}$]{%
		\label{fig:ex3-b}%
		\includegraphics[height=1.6in]{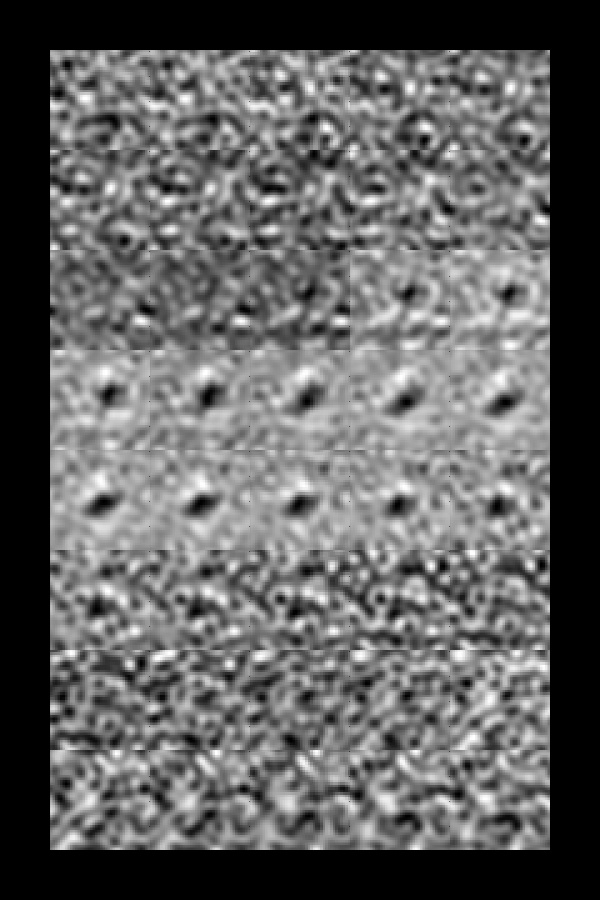}}
	\subfloat[][$C_{2}$]{%
		\label{fig:ex3-c}%
		\includegraphics[height=1.6in]{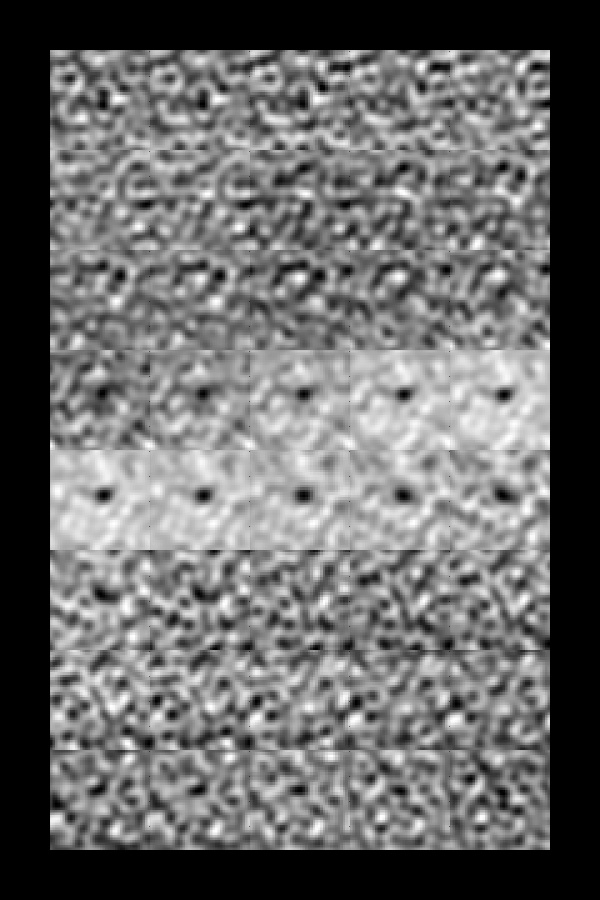}}%
	\hspace{2pt}%
	\subfloat[][$C_{3}$]{%
		\label{fig:ex3-d}%
		\includegraphics[height=1.6in]{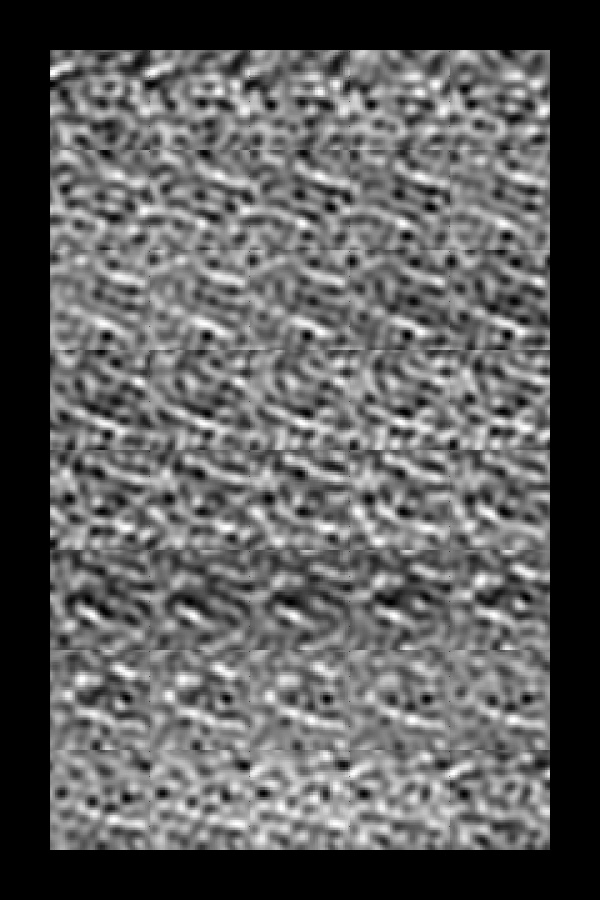}}%

	\caption[]{SNR: 0.547, MWA: 25, Dz: -12, Cs: 2.25 (Simulation Hyperparameters)}
\end{figure}

\begin{figure}
	\centering
	\subfloat[][$C_{0}$]{%
		\label{fig:ex3-a}%
		\includegraphics[height=1.6in]{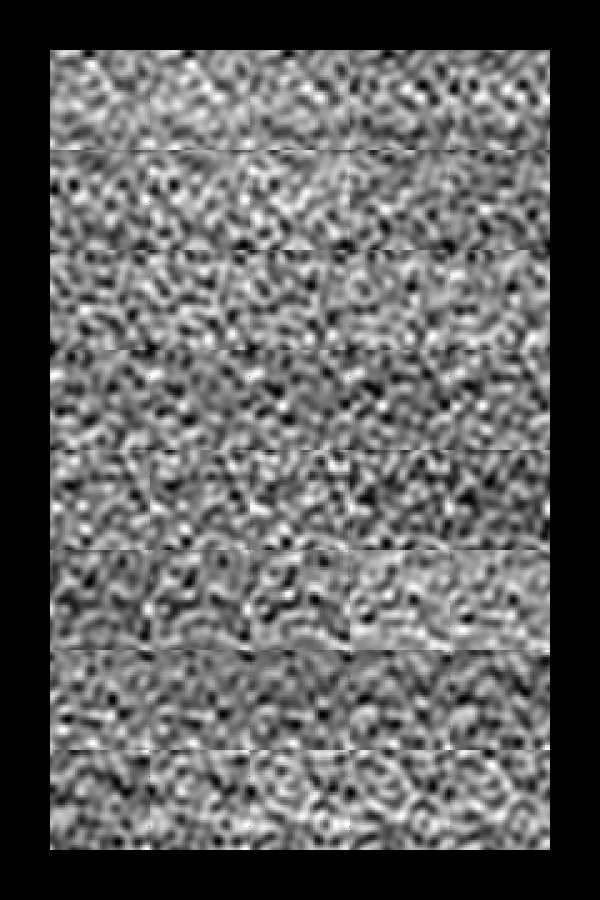}}%
	\hspace{2pt}%
	\subfloat[][$C_{1}$]{%
		\label{fig:ex3-b}%
		\includegraphics[height=1.6in]{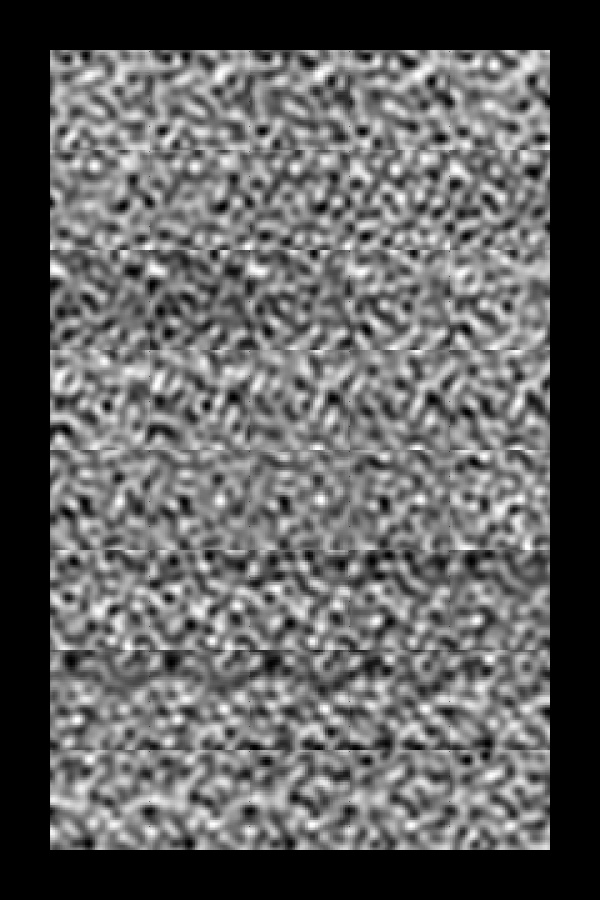}}
	\subfloat[][$C_{2}$]{%
		\label{fig:ex3-c}%
		\includegraphics[height=1.6in]{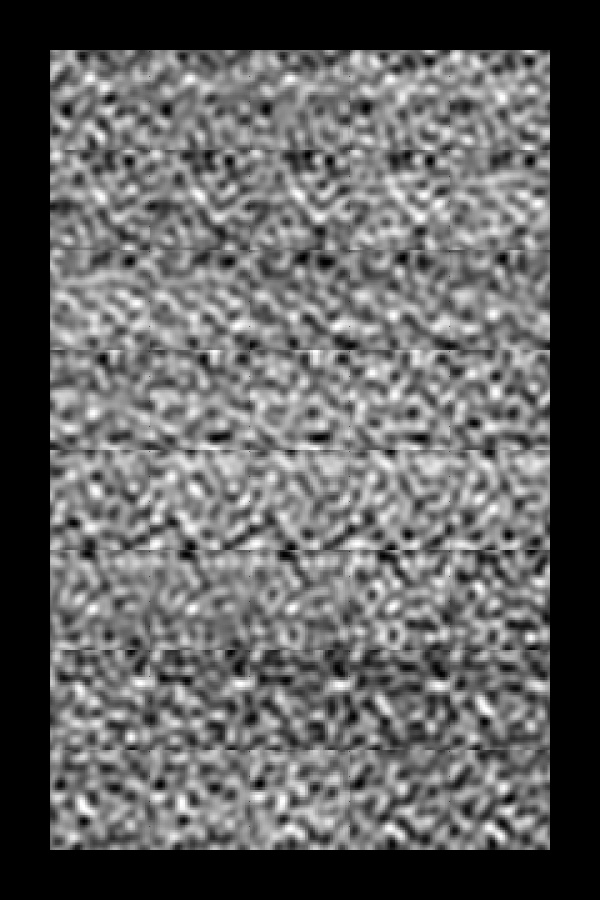}}%
	\hspace{2pt}%
	\subfloat[][$C_{3}$]{%
		\label{fig:ex3-d}%
		\includegraphics[height=1.6in]{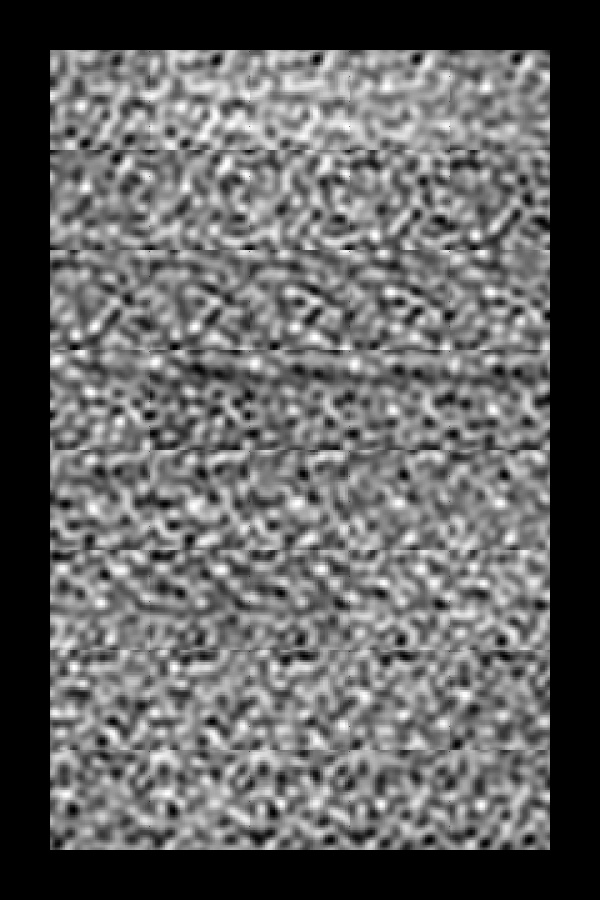}}%

	\caption[]{SNR: 0.547, MWA: 25, Dz: 0, Cs: 2.25 (Simulation Hyperparameters)}
\end{figure}



\end{document}